\documentclass[aip,jcp,reprint,showkeys]{revtex4-1}

\usepackage{graphicx,dcolumn,bm,xcolor,microtype,hyperref,multirow,amsmath,amssymb,amsfonts,physics,mathpazo,float,ulem}
\usepackage[version=4]{mhchem}

\newcommand{\mc}{\multicolumn}
\newcommand{\mcc}[1]{\multicolumn{1}{c}{#1}}

\newcommand{\FeS}{\ce{FeS}}
\newcommand{\DMC}{FN-DMC}
\newcommand{\EDMC}{E_\text{DMC}}
\newcommand{\EFCI}{E_\text{FCI}}
\newcommand{\EexFCI}{E_\text{exFCI}}
\newcommand{\ECI}{E_\text{sCI}}
\newcommand{\EPT}{E_\text{PT2}}
\newcommand{\Dstate}{$^{5}\Delta$}
\newcommand{\Sstate}{$^{5}\Sigma^+$}
\newcommand{\Do}{D_0}
\newcommand{\rFeS}{r_\text{\FeS}}
\newcommand{\re}{r_\text{e}}
\newcommand{\we}{\omega_\text{e}}
\newcommand{\hH}{\Hat{H}}
\newcommand{\PsiSJ}{\Psi_\text{T}^\text{SJ}}
\newcommand{\PsiT}{\Psi_\text{T}}
\newcommand{\PsiCI}{\Psi_\text{sCI}}
\newcommand{\PsiDet}{\Psi_\text{det}}
\newcommand{\Js}{J}
\newcommand{\bR}{\bm{R}}
\newcommand{\bRUp}{\bR^{\uparrow}}
\newcommand{\bRDw}{\bR^{\downarrow}}
\newcommand{\br}{\bm{r}}
\newcommand{\Ndet}{N_\text{det}}
\newcommand{\NdetUp}{N_\text{det}^{\uparrow}}
\newcommand{\NdetDw}{N_\text{det}^{\downarrow}}

\newcommand{\sDetUp}[1]{\mathcal{D}_{#1}^{\uparrow}}
\newcommand{\sDetDw}[1]{\mathcal{D}_{#1}^{\downarrow}}
\newcommand{\DetUp}[1]{D_{#1}^{\uparrow}}
\newcommand{\DetDw}[1]{D_{#1}^{\downarrow}}
\newcommand{\Det}[1]{D_{#1}}
\newcommand{\meh}{m$E_\text{h}$}
\newcommand{\NormUp}[1]{\mathcal{N}_{#1}^{\uparrow}}
\newcommand{\NormDw}[1]{\mathcal{N}_{#1}^{\downarrow}}
\newcommand{\Norm}{\mathcal{N}}

\begin{document}	

\title{Deterministic construction of nodal surfaces within quantum Monte Carlo: the case of {\FeS}}

\author{Anthony Scemama}
\email[Corresponding author: ]{scemama@irsamc.ups-tlse.fr}
\affiliation{Laboratoire de Chimie et Physique Quantiques, Universit\'e de Toulouse, CNRS, UPS, France}
\author{Yann Garniron}
\affiliation{Laboratoire de Chimie et Physique Quantiques, Universit\'e de Toulouse, CNRS, UPS, France}
\author{Michel Caffarel}
\affiliation{Laboratoire de Chimie et Physique Quantiques, Universit\'e de Toulouse, CNRS, UPS, France}
\author{Pierre-Fran{\c c}ois Loos}
\email[Corresponding author: ]{loos@irsamc.ups-tlse.fr}
\affiliation{Laboratoire de Chimie et Physique Quantiques, Universit\'e de Toulouse, CNRS, UPS, France}

\begin{abstract}
In diffusion Monte Carlo (DMC) methods, the nodes (or zeroes) of the trial wave function dictate the magnitude of the fixed-node (FN) error. 
Within standard DMC implementations, they emanate from short multideterminant expansions, \textit{stochastically} optimized in the presence of a Jastrow factor.
Here, following a recent proposal, we follow an alternative route by considering the nodes of selected configuration interaction (sCI) expansions built with the CIPSI (Configuration Interaction using a Perturbative Selection made Iteratively) algorithm. 
In contrast to standard implementations, these nodes can be \textit{systematically} and \textit{deterministically} improved by increasing the size of the sCI expansion.
The present methodology is used to investigate the properties of the transition metal sulfide molecule {\FeS}.
This apparently simple molecule has been shown to be particularly challenging for electronic structure theory methods due to the proximity of two low-energy quintet electronic states of different spatial symmetry.
In particular, we show that, at the triple-zeta basis set level, all sCI results --- including those extrapolated at the full CI (FCI) limit --- disagree with experiment, yielding an electronic ground state of {\Sstate} symmetry.
Performing {\DMC} simulation with sCI nodes, we show that the correct {\Dstate} ground state is obtained if sufficiently large expansions are used.
Moreover, we show that one can systematically get accurate potential energy surfaces and reproduce the experimental dissociation energy as well as other spectroscopic constants.
\end{abstract}


\maketitle

\section{Introduction}
From an experimental point of view, transition metal sulfides have proven to be useful in a variety of fields including biological chemistry, \cite{Crack_2014} catalysis, \cite{Stiefel_1996} and electrochemistry. \cite{Xiao_2016}
From the computational side, the apparently simple {\FeS} diatomic molecule has been giving nightmares to computational chemists.
{The challenging features of the electronic structure of {\FeS} originate from the energetic proximity of two electronic states
\begin{align*}
	\text{\Dstate: } & \sigma^2\pi^4\sigma^2\delta^3\sigma^1\pi^2,
	& 
	\text{\Sstate: }  & \sigma^2\pi^4\sigma^2\delta^2\sigma^2\pi^2,
\end{align*}
with same multiplicity that compete for being the ground state.}
To make things worse, the equilibrium bond lengths associated with these two states are extremely close to each other. 

Experimentally, the ground state of {\FeS} is assigned to be {\Dstate}, \cite{Zhang_1996, Takano_2004} with an equilibrium bond length $\re = 2.017$~\AA, \cite{Takano_2004} and a dissociation energy $\Do = 3.31(15)$ eV. \cite{Drowart_1967}
For this state, the harmonic frequency $\we$ has been estimated to $518 \pm 5$ cm$^{-1}$. \cite{Wang_2011}
Very recently, a much more accurate value of the dissociation energy $\Do = 3.240(3)$ eV has been obtained by Matthew et al.~using predissociation threshold technique. \cite{Matthew_2017}

{\FeS} has been extensively studied by density-functional theory (DFT) and post-Hartree-Fock methods. 
In short, most (but not all) DFT functionals correctly predict a \Dstate~ground state, \cite{Bridgeman_2000,Li_2013, Liang_2009, Schultz_2005, Wu_2007} while
CAS-based multireference methods such as CASSCF/ACPF, \cite{Hubner_1998} CASPT2,\cite{Clima_2007} or CASSCF/ICACPF \cite{Bauschlicher_1995} systematically predict {\Sstate} lower than {\Dstate}.

Here, we investigate this problem using quantum Monte Carlo (QMC).
In recent years, QMC has been applied with great success to a large variety of main group compounds (see e.g.~\onlinecite{Chen_2016, Dubeck__2016, Zhou_2017, Filippi_2016} for recent applications).
Transition metal systems are more challenging but a number of successful studies have also been reported. \cite{Christiansen_1991, Mitas_1993, Belohorec_93, Mitas_1994, Sokolova_2000, Wagner_2003, Diedrich_2005, Caffarel_2005, Buendia_2006, Wagner_2007, Bande_2008, Casula_2009, Caffarel_2010, Petz_2011, Buendia_2013, Caffarel_2014, Scemama_2014, Trail_2015, Doblhoff_Dier_2016, Krogel_2016, Haghighi_Mood_2017}

When multireference effects are weak, QMC is seen as a very accurate method providing benchmark results of a quality similar or superior to the gold-standard CCSD(T).
However, when multireference effects are dominant --- as it is usually the case for metallic compounds with partially-filled $d$ shells --- the situation is more complicated.
Indeed, the results may depend significantly on the trial wave function $\PsiT$ used to guide the walkers through configuration space.
In theory, QMC results should be independent of the choice of $\PsiT$. 
However, it is not true in practice because of the fixed-node approximation which imposes the Schr\"odinger equation to be solved with the additional constraint that the solution vanishes at the zeroes (nodes) of the trial wave function.
Using an approximate $\PsiT$ leads to approximate nodes and, thus, to an approximate energy, known as the FN energy. 
The FN energy being an upper bound of the exact energy, this gives us a convenient variational principle for characterizing the nodal quality: \textit{``the lower the FN energy, the better the nodes''}.
In situations where multireference effects are strong, getting accurate nodes may be difficult.
As we shall see, this is the main challenge we are facing in the present work.

Most QMC studies for transition metal-containing systems have been performed with pseudopotentials.
In this case, an additional source of error, the so-called localization error, is introduced.
This error, specific to QMC, adds up to the standard error associated with the approximate nature of pseudopotentials. 
Similarly to the FN error, the localization error depends on $\PsiT$ and vanishes only for the exact wave function.
Therefore, to get accurate and reliable QMC results, both sources of error have to be understood and controlled.

In 2011, Petz and L\"uchow reported a FN diffusion Monte Carlo ({\DMC}) study of the energetics of diatomic transition metal sulfides from \ce{ScS} to {\FeS} using pseudopotentials and single-determinant trial wave functions. \cite{Petz_2011}
The pseudopotential dependence was carefully investigated, and comparisons with both DFT and CCSD(T) as well as experimental data were performed.
In short, it was found that {\DMC} shows a higher overall accuracy than both B3LYP and CCSD(T) for all diatomics except for \ce{CrS} and {\FeS} that appeared to be particularly challenging.

Very recently, Haghighi-Mood and L\"uchow had a second look at the difficult case of {\FeS}. \cite{Haghighi_Mood_2017} 
In particular, they explored the impact of the level of optimization on the parameters of multideterminant trial wave functions (partial or full optimization of the Jastrow, determinant coefficients and molecular orbitals) on both the FN and localization errors.
Their main conclusions can be summarized as follows.
Using a single-determinant trial wave function made of B3LYP orbitals or fully-optimized orbitals in the presence of a Jastrow factor is sufficient to yield the correct state ordering.
However, in both cases, the dissociation energy is far from the experimental value and thus multideterminant trial wave functions must be employed.
Although a natural choice would be to take into account the missing static correlation via a CASSCF-based trial wave function, they showed that it is insufficient and that a full optimization is essential to get both the correct electronic ground state and reasonable estimates of the spectroscopic constants.

In the present study, we revisit this problem within the original QMC protocol developed in our group these last few years. \cite{Giner_2013, Scemama_2014, Giner_2015, Scemama_2016, Caffarel_2016, Caffarel_2016b, arxiv}
In the conventional protocol, prevailing in the QMC community and employed by Haghighi-Mood and L\"uchow, the nodes of the Slater-Jastrow (SJ) trial wave function 
\begin{equation}
	\label{eq:psiSJ}
	\PsiSJ = \exp(\Js) \, \PsiDet
\end{equation}
are obtained by partially- or fully-optimizing the Jastrow factor $\Js$ and the multiderminant expansion $\PsiDet$ (containing typically a few hundreds or thousands of determinants).
This step is performed in a preliminary variational Monte Carlo calculation by minimizing the energy, the variance of the local energy (or a combination of both) employing one of the optimization methods developed within the QMC context. \cite{Umrigar_2005, Toulouse_2007, Umrigar_2007, Toulouse_2008}
We note that, in practice, the optimization must be carefully monitored because of the large number of parameters (several hundreds or thousands), the nonlinear nature of most parameters (several minima may appear) and the inherent presence of noise in the function to be minimized.

Within our protocol, we rely on configuration interaction (CI) expansions in order to get accurate nodal surfaces, without resorting to the stochastic optimization step.
Our fundamental motivation is to take advantage of all the machinery and experience developed these last decades in the field of wave function methods.
In contrast to the standard protocol described above, the CI nodes can be improved \textit{deterministically} and \textit{systematically} by increasing the size of the CI expansion.
In the present work, we do not introduce any Jastrow factor, essentially to avoid the expensive numerical quadrature involved in the calculation of the pseudopotential, and to facilitate the control of the localization error.
To keep the size of the CI expansion reasonable and retain only the most important determinants, we propose to use selected CI (sCI) algorithms, such as CIPSI (Configuration Interaction using a Perturbative Selection made Iteratively). \cite{Giner_2013}
Using a recently-proposed algorithm to handle large numbers of determinants in \DMC{} \cite{Scemama_2016} we are able to consider up to a few million determinants in our simulations. 

Over the last few years, we have witnessed a re-birth of sCI methods. \cite{Bender_1969, Whitten_1969, Huron_1973, Evangelisti_1983, Cimiraglia_1985, Cimiraglia_1987, Illas_1988, Povill_1992, Abrams_2005, Bunge_2006, Bytautas_2009, Booth_2009, Giner_2013, Knowles_2015, Giner_2015, Caffarel_2014, Scemama_2014, Caffarel_2016, Scemama_2016, Garniron_2017b, Evangelista_2014, Liu_2016, Schriber_2016, Tubman_2016, Holmes_2016, Per_2017, Ohtsuka_2017, Sharma_2017, Holmes_2017, Zimmerman_2017} 
Although these various approaches appear under diverse acronyms, most of them rely on the very same idea of selecting determinants iteratively according to their contribution to the wave function or energy, an idea that goes back to 1969 in the pioneering works of Bender and Davidson, \cite{Bender_1969} and Whitten and Hackmeyer. \cite{Whitten_1969} 
Importantly, we note that any sCI variants can be employed here.

The price to pay for using sCI expansions instead of optimized SJ trial wave functions is the need of much larger multideterminant expansions as well as the presence of larger statistical and systematic errors (such as time-step and basis set incompleteness errors).
However, these disadvantages are compensated by the appealing features of sCI nodes:
i) they are built in a fully-automated way;
ii) they are unique and reproducible;
iii) they can be systematically improved by increasing the level of selection and/or the basis set
(with the possibility of complete basis set extrapolation \cite{Caffarel_2016}), and
iv) they easily produce smooth potential energy surfaces. \cite{Giner_2015}
Unless otherwise stated, atomic units are used throughout.

\begin{table}
	\caption{
		\label{tab:PsiT}
		Characteristics of the various sCI expansions at $\rFeS = 2.0$~{\AA} for various levels of truncation.
		The characteristics of the extrapolated FCI (exFCI) expansion are also reported.
	}
	\begin{ruledtabular}
		\begin{tabular}{lcrrrc}
			\mcc{Method}	&	$\epsilon$		&	\mcc{$\Ndet$}		&	\mcc{$\NdetUp$}	&	\mcc{$\NdetDw$}	&	acronym		\\
			\hline

			sCI			&	$10^{-4}$		&	$15\,723$			&	$191$			&	$188	$			&	sCI(4)		 \\
						&	$10^{-5}$		&	$269\,393$		&	$986	$			&	$1\,191$			&	sCI(5)		\\
						&	$10^{-6}$		&	$1\,127\,071$		&	$3\,883$			&	$4\,623$			&	sCI(6)			\\
						&	$0$			&	$8\,388\,608$		&	$364\,365$		&	$308\,072$		&	sCI($\infty$)			\\
			exFCI		&	\textemdash	&	$\sim 10^{27}$		&	$\sim 10^{16}$		&	$\sim 10^{11}$		&	exFCI			\\
		\end{tabular}
	\end{ruledtabular}
\end{table}

\section{Computational details}
All trial wave functions have been generated with the electronic structure software \textsc{quantum package}, \cite{QP} while the QMC calculations have been performed with the \textsc{qmc=chem} suite of programs. \cite{qmcchem, Scemama_2013} 
Both softwares are developed in our laboratory and are freely available.
For all calculations, we have used the triple-zeta basis sets of Burkatzki et al.\cite{Burkatzki_2007, Burkatzki_2008} (VTZ-ANO-BFD for \ce{Fe} and VTZ-BFD for \ce{S}) in conjunction with the corresponding Burkatzki-Filippi-Dolg (BFD) small-core pseudopotentials including scalar relativistic effects. 
For more details about our implementation of pseudopotentials within QMC, we refer the interested readers to Ref.~\onlinecite{Caffarel_2016b}.
As pointed out by Hammond and coworkers, \cite{Hammond_1987} thanks to the absence of Jastrow factor in our trial wave functions, the non-local pseudopotential can be localized analytically and the usual numerical quadrature over the angular part of the non-local pseudopotential can be eschewed. 
In practice, the calculation of the localized part of the pseudopotential represents only a small overhead (about 15\%) with respect to a calculation without pseudopotential (and the same number of electrons).

In order to compare our results for the dissociation energy of {\FeS} with the experimental value of Matthew et al. \cite{Matthew_2017} and the (theoretical) benchmark value of Haghighi-Mood and L\"uchow, \cite{Haghighi_Mood_2017} we have taken into account the zero-point energy (ZPE) correction, the spin-orbit effects as well as the core-valence correlation contribution the same way as Ref.~\onlinecite{Haghighi_Mood_2017}.
For the {\Dstate} state, this corresponds to an increase of the dissociation energy by $0.06$~eV, and a $0.02$~eV stabilization of {\Dstate} compared to {\Sstate}.

\subsection{
\label{sec:PsiT}
Jastrow-free trial wave functions}
Within the spin-free formalism used in QMC, a CI-based  trial wave function is written as
\begin{equation}
\label{eq:PsiT1}
	\PsiT(\bR)= \sum_{I=1}^{\Ndet} c_I D_{I}(\bR) = \sum_{I=1}^{\Ndet} c_I \DetUp{I}(\bRUp) \DetDw{I}(\bRDw),
\end{equation}
where $\bR=(\br_1,...,\br_N)$ denotes the full set of electronic spatial coordinates, 
$\bRUp$ and $\bRDw$ are the two subsets of spin-up ($\uparrow$) and 
spin-down ($\downarrow$) electronic coordinates, and $D_I^{\sigma}(\bR^{\sigma})$ ($\sigma =$ $\uparrow \text{ or } \downarrow$) are spin-specific determinants.

In practice, the various products $\DetUp{I} \DetDw{I}$ contain many 
identical spin-specific determinants. 
For computational efficiency, it is then advantageous to group them and compute only once their contribution to the wave function and its derivatives. \cite{Scemama_2016}
Therefore, the Jastrow-free CI trial wave functions employed in the present study are rewritten 
in a ``spin-resolved'' form
\begin{equation}
	\label{eq:PsiT2}
	\PsiT(\bR) = \sum_{i=1}^{\NdetUp} \sum_{j=1}^{\NdetDw} c_{ij} \sDetUp{i}(\bRUp) \sDetDw{j}(\bRDw),
\end{equation}
where $\qty{\mathcal{D}^{\sigma}_i}_{i=1,\ldots,\Ndet^\sigma}$ denotes the set of all \textit{distinct} spin-specific determinant appearing in Eq.~\eqref{eq:PsiT1}.

\subsection{Quantum Monte Carlo calculations}

To avoid handling too many determinants in $\PsiT$, a truncation scheme has to be introduced. 
In most CI and/or QMC calculations, the expansion is truncated by either introducing a cutoff on the CI coefficients or on the norm of the wave function. 
Here, we use an alternative truncation scheme knowing that most of the computational effort lies in the calculation of the spin-specific determinants and their derivatives. 
Removing a product of determinants whose spin-specific determinants are already present in other products does not change significantly the computational cost. Accordingly, a natural choice is then 
to truncate the wave function by removing \textit{independently} spin-up and spin-down determinants.
To do so, we decompose the norm of the wave function as
\begin{equation}
	\Norm = \sum_{i=1}^{\NdetUp} \sum_{j=1}^{\NdetDw} \abs{c_{ij}}^2
	= \sum_{i=1}^{\NdetUp} \NormUp{i} 
	= \sum_{j=1}^{\NdetDw} \NormDw{j}.
\end{equation}
A determinant $\sDetUp{i}$ is retained in $\PsiT$ if 
\begin{equation}
	\label{eq:trunc}
	\NormUp{i} = \sum_{j=1}^{\NdetDw} \abs{c_{ij}}^2 > \epsilon,
\end{equation}
where $\epsilon$ is a user-defined threshold. 
A similar formula is used for $\sDetDw{j}$.
When $\epsilon = 0$, the entire set of determinants is retained in the QMC simulation.

In order to treat the two electronic states ({\Sstate} and {\Dstate}) on equal footing, a common set of spin-specific determinants $\qty{\mathcal{D}^{\sigma}_i}_{i=1,\ldots,\Ndet^\sigma}$ is used for both states.
In addition, a common set of molecular orbitals issued from a preliminary state-averaged CASSCF calculation is employed.  
These CASSCF calculations have been performed with the GAMESS package \cite{Schmidt_1993} while, for the atoms, we have performed ROHF calculations.
The active space contains 12 electrons and 9 orbitals ($3d$ and $4s$ orbitals of \ce{Fe} and $3p$ orbitals of \ce{S}).
The multideterminant expansion \eqref{eq:PsiT1} has been constructed using the sCI algorithm CIPSI, \cite{Huron_1973, Evangelisti_1983} which uses a second-order perturbative criterion to select the energetically-important determinants $\Det{I}$ in the FCI space. \cite{Giner_2013, Giner_2015, Caffarel_2014, Scemama_2014, Caffarel_2016, Scemama_2016, Garniron_2017b}
A $n_s$-state truncated sCI expansion (here $n_s=2$) is obtained via a natural generalization of the state-specific criterion introduced in Eq.~\eqref{eq:trunc}:
a determinant $\sDetUp{i}$ is retained in $\PsiT$ if 
\begin{equation}
	\NormUp{i} = \frac{1}{n_s} \sum_{k=1}^{n_s} \sum_{j=1}^{\NdetDw} \abs{c^{(k)}_{ij}}^2 > \epsilon,
\end{equation}
with a similar formula for $\sDetDw{j}$.

The characteristics of the various trial wave functions considered here (and their acronyms) at $\rFeS = 2.0$~{\AA} are presented in Table \ref{tab:PsiT}. 
For other $\rFeS$ values, the numbers of determinants are slightly different.
Our largest sCI trial wave function contains $8\,388\,608$ determinants and is labeled sCI($\infty$). 
The sCI($n$) wave functions with $n=4$, 5, and 6 are obtained by truncation of the sCI($\infty$) expansion setting $\epsilon=10^{-n}$. 
They contain respectively $15\,723$, $269\,393$, and $1\,127\,071$ determinants.
At this stage, we are not able to use the entire $8\,388\,608$ determinants of the sCI($\infty$) wave function within our {\DMC} simulations.
In comparison, Haghighi-Mood and L\"uchow's CASSCF-based trial wave function (labelled as HML in Table \ref{tab:BigTab}) only contains 630 and 500 determinants for the {\Sstate} and {\Dstate} states, respectively. \cite{Haghighi_Mood_2017}
However, as discussed in the introduction, fully-optimized SJ trial wave functions require much smaller multireference expansions.

Based on these trial wave functions, we performed {\DMC} calculations
with the stochastic reconfiguration algorithm developed by Assaraf et al. \cite{Assaraf_2000}
In order to remove the time-step error, all our {\DMC} results have been extrapolated to zero time-step using a two-point linear extrapolation with $\tau = 2 \times 10^{-4}$ and $4 \times 10^{-4}$. \cite{Lee_2011}
Note that, because the variance of the local energy is larger than in SJ calculations, time-step errors are enhanced and shorter time steps are required.

\begin{table*}
        \caption{
                \label{tab:BigTab}
                \DMC{} energies $\EDMC$ (in Hartrees) at equilibrium geometry, dissociation energy $\Do$ (in eV), equilibrium distance $\re$ (in \AA), harmonic frequency $\we$ (in cm$^{-1}$) for the \Sstate~and \Dstate~of {\FeS} obtained with various trial wave functions $\PsiT$.
                The error bar corresponding to one standard error is reported in parenthesis.
        }
        \begin{ruledtabular}
                \begin{tabular}{lcccccccccc}
                $\PsiT$         &       \mc{3}{c}{\FeS~(\Sstate)}       &       \mc{3}{c}{\FeS~(\Dstate)}
                                        &       \ce{Fe}~($^5$D) &       \ce{S}~($^3$P)  &        $\Do$          &       Ref.    \\
                                \cline{2-4}     \cline{5-7}     \cline{8-8}     \cline{9-9}
                                        &       $\EDMC$                 &       $\re$           &       $\we$
                                        &       $\EDMC$                 &       $\re$           &       $\we$
                                        &       $\EDMC$                 &       $\EDMC$         \\
                 \hline
                HML                     &       $-134.057\,1(4)$     &       $2.00(1)$               &       $518(7)$ 
                                        &       $-134.057\,9(4)$     &       $2.031(7)$      &       $499(11)$
                                        &       $-123.812\,6(4)$     &       $-10.131\,4(1)$      &       $3.159(15)$     &       \onlinecite{Haghighi_Mood_2017}                 \\
                sCI(4)          &       $-134.010\,1(8)$        &       $1.994(7)$              &       $532(20)$
                                        &       $-134.004\,0(7)$        &       $2.029(7)$              &       $502(15)$
                                        &       $-123.802\,8(9)$     &       $-10.127\,9(2)$      &       $2.055(20)$                     &       This work                                       \\
                sCI(5)          &       $-134.047\,9(10)$       &       $1.992(8)$              &       $551(24)$
                                        &       $-134.040\,2(10)$       &       $2.048(11)$             &       $489(21)$
                                        &       $-123.823\,4(10)$       &       $-10.131\,2(2)$      &       $2.389(28)$                     &       This work                                       \\
                sCI(6)          &       $-134.061\,7(14)$       &       $1.994(12)$             &       $497(35)$
                                        &       $-134.067\,1(14)$       &       $2.004(11)$             &       $550(32)$
                                        &       $-123.830\,0(12)$       &       $-10.133\,4(3)$      &       $3.062(39)$                     &       This work                                       \\
                exFCI               &       $-134.086\,3(15)$       &       $1.990(12)$             &       $523(37)$
                                        &       $-134.088\,5(18)$       &       $2.016(14)$             &       $525(40)$
                                        &       $-123.837\,2(12)$       &       $-10.133\,6(3)$      &       $3.267(49)$                     &       This work                                       \\
                Exp.                    &       \textemdash             &       \textemdash     &       \textemdash
                                        &       \textemdash             &       $2.017$         &       $518(5)$
                                        &       \textemdash             &       \textemdash     &       $3.240(3)$   &       \onlinecite{Takano_2004, Wang_2011, Matthew_2017}\\
                \end{tabular}
        \end{ruledtabular}
\end{table*}

\begin{figure}
	\includegraphics[width=\linewidth]{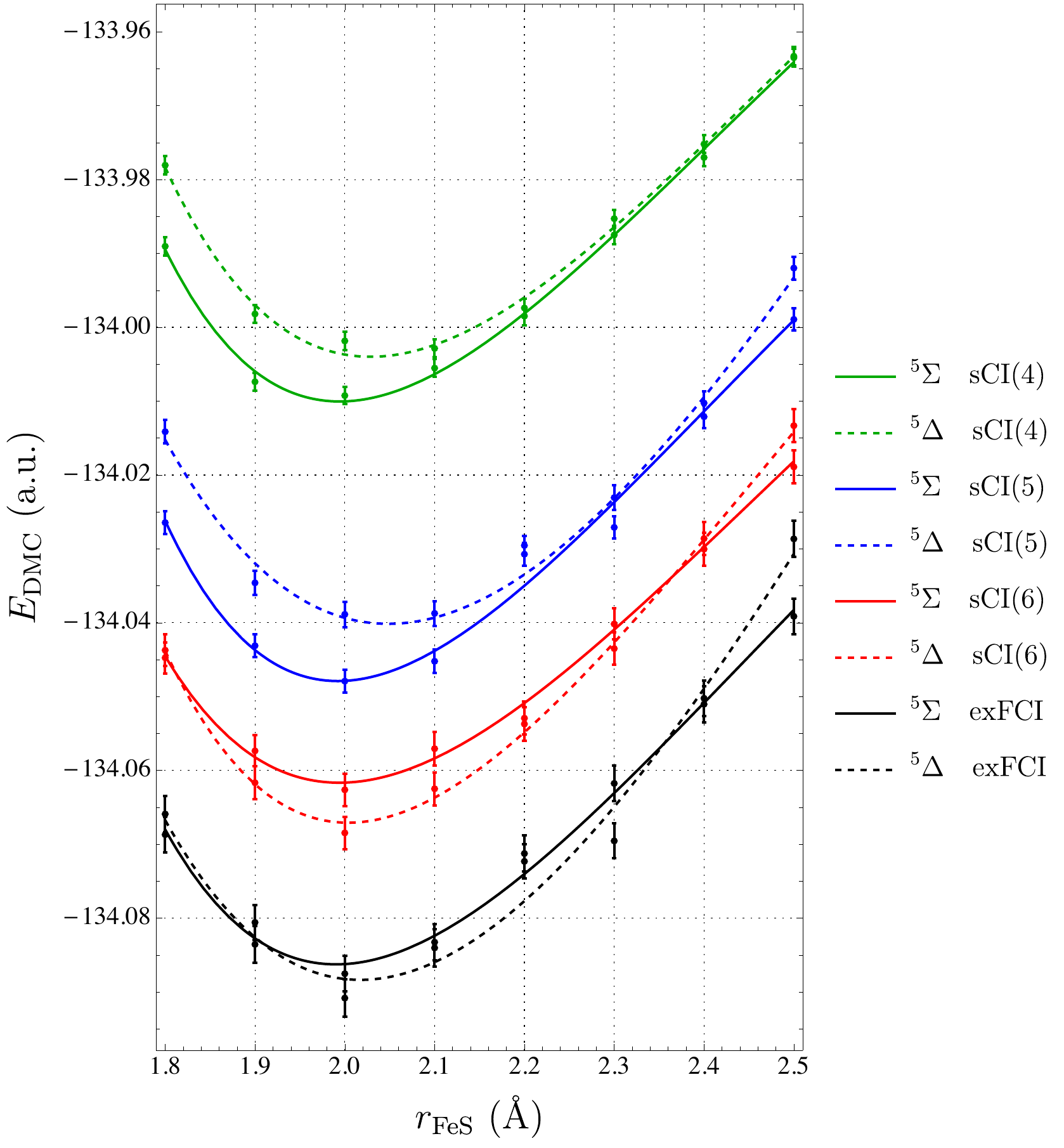}
	\caption{
		\label{fig:DMC_curve}
		$\EDMC$ (in Hartrees) for the {\Sstate} (solid) and {\Dstate} (dashed) states of {\FeS} as a function of $\rFeS$ (in \AA) for various trial wave functions.
		The error bar corresponds to one standard error.
	}
\end{figure}

\begin{figure}
	\includegraphics[width=\linewidth]{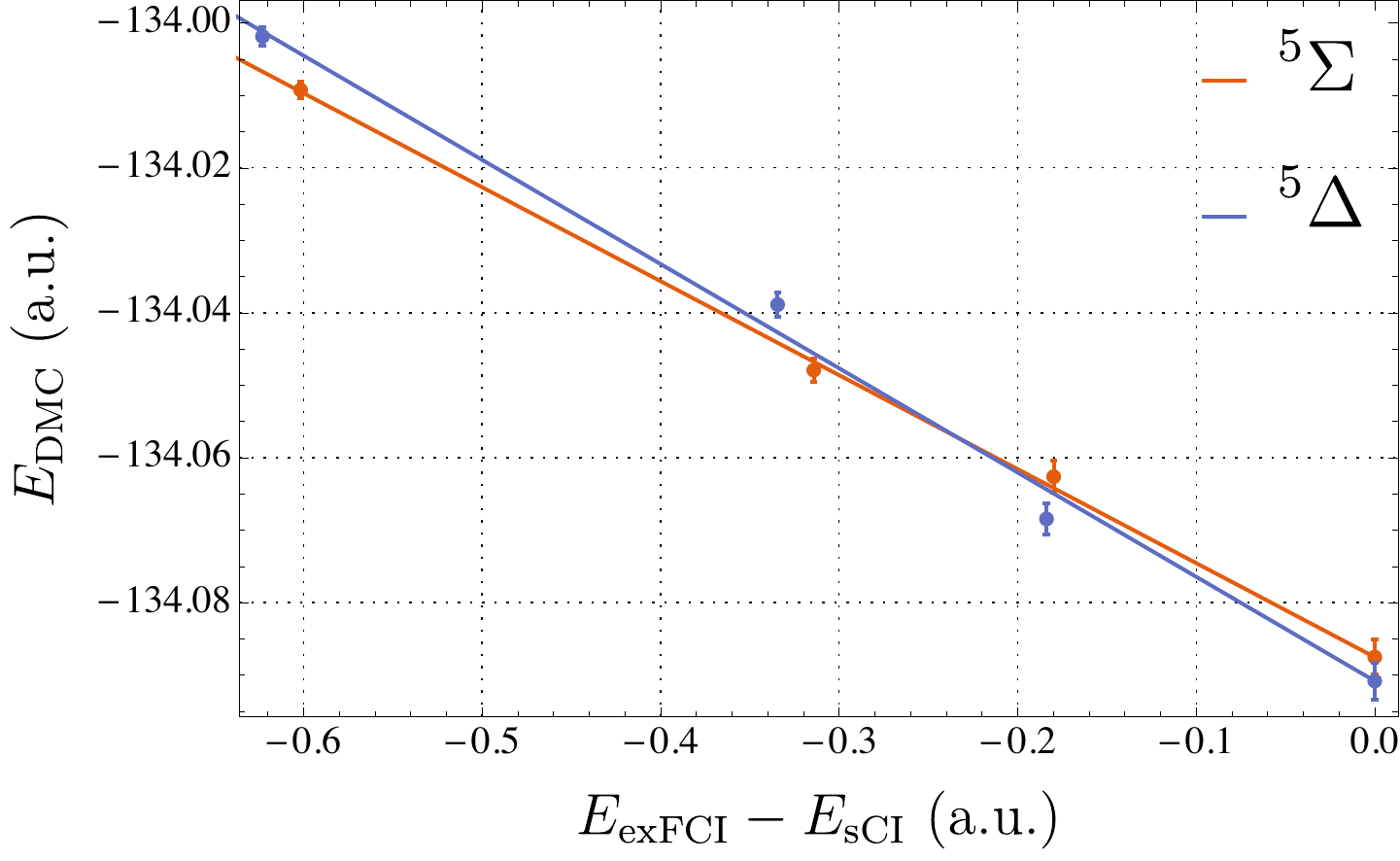}
	\caption{
		\label{fig:extrap_FCI}
		Three-point linear extrapolation of the {\DMC} energy $\EDMC$ to the FCI limit ($\EexFCI - \ECI =0$) for the {\Sstate} (red) and {\Dstate} (blue) states of {\FeS} at $\rFeS = 2.0$ \AA.
		The error bar corresponds to one standard error.
	}
\end{figure}

\begin{table*}
	\caption{
		\label{tab:BigTabAt}
		sCI energy $\ECI$, second-order perturbation correction $\EPT$ and {\DMC} energy $\EDMC$ (in Hartrees) for the \ce{Fe} ($^5$D state) and \ce{S} ($^3$P state) atoms obtained with various methods.
		The error bar corresponding to one standard error is reported in parenthesis.
	}
	\begin{ruledtabular}
		\begin{tabular}{lcccccccc}
		Method			&	\mc{3}{c}{\ce{Fe}~($^5$D)}								&	\mc{3}{c}{\ce{S}~($^3$P)}			\\
						\cline{2-4}												\cline{5-7}
						&	$\ECI$			&	$\EPT$		&	$\EDMC$			&	$\ECI$			&	$\EPT$		&	$\EDMC$			\\	
		\hline
		sCI(4)			&	$-123.418\,124$	&	$-0.316\,44(3)$		&	$-123.802\,8(9)$	&	$-10.093\,850$	&	$-0.023\,348(2)$	&	$-10.127\,9(2)$	\\
		sCI(5)			&	$-123.607\,608$	&	$-0.127\,40(1)$		&	$-123.823\,4(10)$	&	$-10.108\,576$	&	$-0.007\,941(1)$	&	$-10.131\,2(2)$	\\
		sCI(6)        		&	$-123.673\,435$	&	$-0.063\,698(6)$	&	$-123.830\,0(12)$	&	$-10.113\,926$	&	$-0.002\,179(0)$	&	$-10.133\,4(3)$	\\
		sCI($\infty$)		&	$-123.720\,629$	&	$-0.016\,987(2)$	&	\textemdash			&	$-10.115\,844$	&	$-0.000\,188(0)$	&	\textemdash		\\
		exFCI			&	$-123.738\,264$	&	$0$					&	$-123.837\,2(12)$	&	$-10.115\,996$	&	$0$					&	$-10.133\,4(3)$	\\
		\end{tabular}
	\end{ruledtabular}
\end{table*}

\begin{table*}
	\caption{
		\label{tab:BigTabMol}
		sCI energy $\ECI$, second-order perturbation correction $\EPT$ and {\DMC} energy $\EDMC$ (in Hartrees) for the \Sstate~and \Dstate~of {\FeS} obtained with various methods.
		The error bar corresponding to one standard error is reported in parenthesis.
	}
	\begin{ruledtabular}
		\begin{tabular}{lccccccccc}
		Method		&	$\rFeS$	&	\mc{3}{c}{\FeS~(\Sstate)}	&	\mc{3}{c}{\FeS~(\Dstate)}			\\
								\cline{3-5}					\cline{6-8}
					&			&	$\ECI$			&	$\EPT$		&	$\EDMC$			&	$\ECI$			&	$\EPT$		&	$\EDMC$			\\	
		\hline
		sCI(4)		&	1.8		&	$-133.322\,421$	&	$-0.602\,63(6)$	&	$-133.989\,0(12)$	&	$-133.287\,542$	&	$-0.633\,28(6)$	&	$-133.978\,1(12)$	\\
					&	1.9		&	$-133.338\,106$	&	$-0.603\,60(6)$	&	$-134.007\,4(12)$	&	$-133.304\,261$	&	$-0.635\,26(6)$	&	$-133.998\,2(12)$	\\
					&	2.0		&	$-133.344\,260$	&	$-0.600\,73(6)$	&	$-134.009\,2(12)$	&	$-133.316\,924$	&	$-0.624\,38(6)$	&	$-134.001\,8(12)$	\\
					&	2.1		&	$-133.343\,984$	&	$-0.597\,12(6)$	&	$-134.005\,5(12)$	&	$-133.318\,747$	&	$-0.617\,18(6)$	&	$-134.002\,9(12)$	\\
					&	2.2		&	$-133.337\,856$	&	$-0.595\,60(6)$	&	$-133.998\,5(12)$	&	$-133.321\,678$	&	$-0.603\,51(6)$	&	$-133.997\,4(12)$	\\
					&	2.3		&	$-133.312\,474$	&	$-0.606\,32(6)$	&	$-133.987\,5(12)$	&	$-133.334\,006$	&	$-0.587\,80(6)$	&	$-133.985\,3(12)$	\\
					&	2.4		&	$-133.309\,214$	&	$-0.601\,37(6)$	&	$-133.977\,0(12)$	&	$-133.323\,583$	&	$-0.587\,32(6)$	&	$-133.975\,2(12)$	\\
					&	2.5		&	$-133.295\,921$	&	$-0.607\,48(6)$	&	$-133.963\,3(13)$	&	$-133.305\,592$	&	$-0.595\,18(6)$	&	$-133.963\,5(12)$	\\
					\hline
		sCI(5)		&	1.8		&	$-133.607\,003$	&	$-0.290\,22(3)$	&	$-134.026\,4(15)$	&	$-133.574\,416$	&	$-0.309\,14(3)$	&	$-134.014\,1(16)$	\\
					&	1.9		&	$-133.623\,662$	&	$-0.290\,03(3)$	&	$-134.043\,1(15)$	&	$-133.595\,588$	&	$-0.305\,97(3)$	&	$-134.034\,6(16)$	\\
					&	2.0		&	$-133.631\,686$	&	$-0.286\,36(3)$	&	$-134.047\,9(16)$	&	$-133.605\,430$	&	$-0.302\,44(3)$	&	$-134.038\,9(17)$	\\
					&	2.1		&	$-133.631\,076$	&	$-0.283\,06(3)$	&	$-134.045\,2(16)$	&	$-133.608\,180$	&	$-0.297\,83(3)$	&	$-134.038\,7(17)$	\\
					&	2.2		&	$-133.622\,872$	&	$-0.283\,00(3)$	&	$-134.034\,0(15)$	&	$-133.600\,172$	&	$-0.299\,37(3)$	&	$-134.029\,5(13)$	\\
					&	2.3		&	$-133.600\,838$	&	$-0.289\,70(3)$	&	$-134.023\,1(17)$	&	$-133.618\,393$	&	$-0.276\,15(3)$	&	$-134.027\,1(15)$	\\
					&	2.4		&	$-133.591\,037$	&	$-0.289\,75(3)$	&	$-134.012\,1(15)$	&	$-133.602\,716$	&	$-0.279\,50(3)$	&	$-134.010\,3(16)$	\\
					&	2.5		&	$-133.583\,356$	&	$-0.286\,68(3)$	&	$-133.998\,9(15)$	&	$-133.590\,900$	&	$-0.278\,56(3)$	&	$-134.992\,0(15)$	\\
					\hline
		sCI(6)		&	1.8		&	$-133.742\,093$	&	$-0.159\,64(2)$	&	$-134.044\,8(21)$	&	$-133.725\,217$	&	$-0.162\,79(2)$	&	$-134.043\,7(21)$	\\
					&	1.9		&	$-133.759\,798$	&	$-0.157\,96(2)$	&	$-134.057\,4(21)$	&	$-133.746\,303$	&	$-0.160\,67(2)$	&	$-134.061\,7(22)$	\\
					&	2.0		&	$-133.766\,027$	&	$-0.155\,98(2)$	&	$-134.062\,6(22)$	&	$-133.756\,008$	&	$-0.157\,98(2)$	&	$-134.068\,5(22)$	\\
					&	2.1		&	$-133.764\,192$	&	$-0.154\,31(2)$	&	$-134.057\,1(23)$	&	$-133.758\,610$	&	$-0.154\,21(2)$	&	$-134.062\,5(22)$	\\
					&	2.2		&	$-133.759\,178$	&	$-0.151\,39(2)$	&	$-134.052\,9(22)$	&	$-133.755\,183$	&	$-0.151\,49(2)$	&	$-134.053\,7(22)$	\\
					&	2.3		&	$-133.749\,068$	&	$-0.148\,46(1)$	&	$-134.040\,2(21)$	&	$-133.750\,644$	&	$-0.148\,52(1)$	&	$-134.043\,5(22)$	\\
					&	2.4		&	$-133.741\,085$	&	$-0.146\,21(1)$	&	$-134.028\,6(23)$	&	$-133.740\,633$	&	$-0.146\,76(1)$	&	$-134.030\,0(22)$	\\
					&	2.5		&	$-133.731\,347$	&	$-0.145\,00(1)$	&	$-134.018\,9(23)$	&	$-133.729\,703$	&	$-0.145\,52(1)$	&	$-134.013\,3(22)$	\\
					\hline
	sCI($\infty$)	&	1.8		&	$-133.850\,093$	&	$-0.071\,305(7)$	&	\textemdash		&	$-133.836\,804$	&	$-0.073\,473(7)$	&	\textemdash		\\
					&	1.9		&	$-133.868\,551$	&	$-0.069\,523(7)$	&	\textemdash		&	$-133.857\,909$	&	$-0.071\,490(7)$	&	\textemdash		\\
					&	2.0		&	$-133.874\,845$	&	$-0.067\,992(7)$	&	\textemdash		&	$-133.866\,607$	&	$-0.069\,933(7)$	&	\textemdash		\\
					&	2.1		&	$-133.873\,424$	&	$-0.066\,144(7)$	&	\textemdash		&	$-133.867\,130$	&	$-0.068\,250(7)$	&	\textemdash		\\
					&	2.2		&	$-133.868\,534$	&	$-0.063\,444(6)$	&	\textemdash		&	$-133.864\,246$	&	$-0.065\,508(7)$	&	\textemdash		\\
					&	2.3		&	$-133.856\,234$	&	$-0.063\,467(6)$	&	\textemdash		&	$-133.858\,730$	&	$-0.061\,578(6)$	&	\textemdash		\\
					&	2.4		&	$-133.849\,753$	&	$-0.058\,745(6)$	&	\textemdash		&	$-133.849\,494$	&	$-0.059\,879(6)$	&	\textemdash		\\
					&	2.5		&	$-133.839\,274$	&	$-0.056\,324(6)$	&	\textemdash		&	$-133.840\,801$	&	$-0.057\,100(6)$	&	\textemdash		\\
					\hline
		exFCI		&	1.8		&	$-133.924\,967$	&	$0$			&	$-134.068\,7(24)$	&	$-133.913\,076$	&	$0$			&	$-134.065\,9(25)$	\\
					&	1.9		&	$-133.941\,175$	&	$0$			&	$-134.080\,6(24)$	&	$-133.932\,524$	&	$0$			&	$-134.083\,6(25)$	\\
					&	2.0		&	$-133.945\,859$	&	$0$			&	$-134.087\,5(24)$	&	$-133.939\,854$	&	$0$			&	$-134.090\,8(26)$	\\
					&	2.1		&	$-133.942\,722$	&	$0$			&	$-134.083\,3(25)$	&	$-133.938\,821$	&	$0$			&	$-134.084\,0(25)$	\\
					&	2.2		&	$-133.934\,789$	&	$0$			&	$-134.074\,4(24)$	&	$-133.932\,709$	&	$0$			&	$-134.072\,3(23)$	\\
					&	2.3		&	$-133.922\,799$	&	$0$			&	$-134.061\,8(24)$	&	$-133.923\,119$	&	$0$			&	$-134.069\,5(24)$	\\
					&	2.4		&	$-133.910\,763$	&	$0$			&	$-134.050\,2(24)$	&	$-133.911\,892$	&	$0$			&	$-134.051\,0(25)$	\\
					&	2.5		&	$-133.897\,709$	&	$0$			&	$-134.039\,1(24)$	&	$-133.900\,371$	&	$0$			&	$-134.028\,7(24)$	\\
		\end{tabular}
	\end{ruledtabular}
\end{table*}

\section{Results and discussion}
In Table \ref{tab:BigTab}, we report {\DMC} energies at equilibrium geometry as well as other quantities of interest such as the dissociation energy $\Do$, the equilibrium distance $\re$ and harmonic frequency $\we$ obtained with various trial wave functions.
These values are obtained via the standard four-parameter Morse potential representation 
of the numerical values gathered in Tables \ref{tab:BigTabAt} and 
\ref{tab:BigTabMol}.\footnote{The error bars have been obtained by fitting a 
large set of energy curves. 
Each of these curves is obtained from independent realizations 
of the statistical noise. Note that due to the absence of correlations in the statistical noise, 
the error bars obtained in this way are certainly overestimated.}
For comparison purposes, Haghighi-Mood and L\"uchow's results are also reported based on their best trial wave function. \cite{Haghighi_Mood_2017} 
When available, the experimental result is also reported. \cite{Takano_2004, Wang_2011, Drowart_1967, Matthew_2017}
The value of $\Do$ is always calculated with respect to the {\Dstate} state adding the corresponding corrections for ZPE, spin-orbit effects and core-valence correlation, as described above (see Sec.~\ref{sec:PsiT}). 
The dissociation profile of {\FeS} obtained with {\DMC} is depicted in Fig.~\ref{fig:DMC_curve} for various trial wave functions.

For the variational results gathered in Tables \ref{tab:BigTabAt} and \ref{tab:BigTabMol}, the FCI limit has been reached by the method recently proposed by Holmes, Umrigar and Sharma \cite{Holmes_2017} in the context of the  (selected) heat-bath CI method. \cite{Holmes_2016, Sharma_2017, Holmes_2017}
In order to obtain FCI results, they proposed to linearly extrapolate the sCI energy $\ECI$ as a function of the second-order Epstein-Nesbet energy 
\begin{equation}
\label{eq:pt2}
	\EPT = \sum_\alpha \frac{\abs{\mel{\alpha}{\hH}{\PsiCI}}^2}{\ECI-\mel{\alpha}{\hH}{\alpha}},
\end{equation}
which is an estimate of the truncation error in the sCI algorithm, i.e $\EPT \approx \EFCI-\ECI$. \cite{Huron_1973}
In Eq.~\eqref{eq:pt2}, the sum runs over all external determinants $\ket{\alpha}$ (i.e.~not belonging to the sCI expansion) connected via $\hH$ to the sCI wave function $\PsiCI$, i.e.~$\mel{\alpha}{\hH}{\PsiCI} \neq 0$.
When $\EPT = 0$, the FCI limit has effectively been reached.
In our case, $\EPT$ is efficiently evaluated thanks to our recently-proposed hybrid stochastic-deterministic algorithm, \cite{Garniron_2017b} which explains the error bar on $\EPT$ in Tables \ref{tab:BigTabAt} and \ref{tab:BigTabMol}.
The extrapolated FCI results are labeled exFCI from hereon.
To obtain the {\DMC} curve with an \textit{effective} FCI trial wave function, we have generalized the extrapolation procedure described above, and we have performed a three-point linear extrapolation of the {\DMC} energy as a function of $\EexFCI-\ECI$ using the sCI(4), sCI(5) and sCI(6) results (see Fig.~\ref{fig:extrap_FCI}).

The first observation we would like to make is that, at the variational level, the {\Dstate} state is never found lower in energy than the {\Sstate} state, even after performing the extrapolation to the FCI limit (see Table \ref{tab:BigTabMol}). 
Because all post-Hartree-Fock methods are indeed an approximation of FCI, they are expected to predict a {\Sstate} ground state for this particular basis set.
This observation is in agreement with the CASPT2 results previously published in the literature. \cite{Bauschlicher_1995, Hubner_1998, Clima_2007}
Thus, one can attribute the wrong state ordering to basis set incompleteness, the only remaining approximation. 

At the {\DMC} level, one must include at least a few hundred thousand determinants in order to find the proper ground state.
For larger $\epsilon$ values ($10^{-4}$ and $10^{-5}$), $\Do$ is underestimated due to the unbalanced treatment of the isolated atoms compared to the dimer at equilibrium geometry.
Indeed, for a given number of determinants, the energy of the atomic species is much closer to the FCI limit than the energy of {\FeS}.

For $\epsilon = 10^{-6}$, our approach correctly predicts a {\Dstate} ground state.
However, although our {\DMC} energies are much lower than those obtained with the HML trial wave function, our estimate of the dissociation energy ($\Do = 3.062(39)$~eV) is still below the experimental value.
This underestimation of $\Do$ can be ultimately tracked down to the lack of size-consistency of the truncated CI wave function. 
With more than $10^{6}$ determinants in the variational space, the wave function is still $150$~\meh~higher than the exFCI wave function, while the atoms are much better described by the sCI wave function.
To remove the size-consistency error, we then extrapolate the {\DMC} energies to the (size-consistent) FCI limit of the trial wave function, as described above.

In that case, using the extrapolated {\DMC} energies of the molecule and isolated atoms reported in Table \ref{tab:BigTab}, we obtain a value of $\Do = 3.267(49)$ eV, which nestles nicely between the experimental values of Matthew et al.\cite{Matthew_2017}~($3.240(3)$ eV) and Drowart et al.\cite{Drowart_1967}~($3.31(15)$ eV).
As a final remark, we note that other spectroscopic constants, such as the equilibrium geometry and the harmonic frequency, are fairly well reproduced by our approach.

\section{Conclusion}
In this article, the potential energy curves of two electronic states --- {\Dstate} and {\Sstate} --- of the {\FeS} molecule have been calculated using the sCI algorithm CIPSI and the stochastic {\DMC} method.
In all our sCI calculations, {\Sstate} is found to be the ground state, in disagreement with experiment.
It is not only true for all CIPSI expansions with up to 8 million determinants but also for the estimated FCI limit obtained using the extrapolation procedure recently proposed by Holmes et al. \cite{Holmes_2017}

This conclusion agrees with other high-level ab initio wave function calculations which all wrongly predict a ground state of {\Sstate} symmetry. 
{\DMC} calculations have been performed using CIPSI expansions including up to $1\,127\,071$ selected determinants as trial wave functions.
Contrary to standard QMC calculations, we do not introduce any Jastrow factor: the CI expansions have been used as they are (no optimization).
It is found that, when the number of determinants in the trial wave function reaches few hundred thousands, the {\DMC} ground state switches from the {\Sstate} state to the correct {\Dstate} state, as predicted experimentally.

Generalizing the extrapolation procedure of Holmes et al., \cite{Holmes_2017} an estimate of the {\DMC} potential energy curves corresponding to the FCI nodes can be obtained. 
The resulting dissociation energy is found to be $3.267(49)$ eV, in agreement with the recent experimental value of Matthew et al.~($3.240(3)$ eV). \cite{Matthew_2017}
As already observed in previous applications, the {\DMC} energy obtained with CIPSI nodes is found to systematically decrease as a function of the number of selected determinants. \cite{Giner_2013, Giner_2015, Caffarel_2014, Scemama_2014, Caffarel_2016, Caffarel_2016b}
For the largest expansion, our fixed-node energies are lower than the values recently reported by Haghighi-Mood and L\"uchow \cite{Haghighi_Mood_2017} using a fully-optimized SJ trial wave function.
This important result illustrates that ``pure'' sCI nodes is a realistic alternative to stochastically-optimized SJ trial wave functions, even for a challenging system such as {\FeS}.
A similar conclusion had already been drawn in our recent study of the water molecule. \cite{Caffarel_2016}

\begin{acknowledgments}
The authors would like to thank Arne L\"uchow for numerous stimulating discussions.
This work was performed using HPC resources from CALMIP (Toulouse) under allocation 2016-0510 and from GENCI-TGCC (Grant 2016-08s015).
\end{acknowledgments}

%


\begin{thebibliography}{85}%
\makeatletter
\providecommand \@ifxundefined [1]{%
 \@ifx{#1\undefined}
}%
\providecommand \@ifnum [1]{%
 \ifnum #1\expandafter \@firstoftwo
 \else \expandafter \@secondoftwo
 \fi
}%
\providecommand \@ifx [1]{%
 \ifx #1\expandafter \@firstoftwo
 \else \expandafter \@secondoftwo
 \fi
}%
\providecommand \natexlab [1]{#1}%
\providecommand \enquote  [1]{``#1''}%
\providecommand \bibnamefont  [1]{#1}%
\providecommand \bibfnamefont [1]{#1}%
\providecommand \citenamefont [1]{#1}%
\providecommand \href@noop [0]{\@secondoftwo}%
\providecommand \href [0]{\begingroup \@sanitize@url \@href}%
\providecommand \@href[1]{\@@startlink{#1}\@@href}%
\providecommand \@@href[1]{\endgroup#1\@@endlink}%
\providecommand \@sanitize@url [0]{\catcode `\\12\catcode `\$12\catcode
  `\&12\catcode `\#12\catcode `\^12\catcode `\_12\catcode `\%12\relax}%
\providecommand \@@startlink[1]{}%
\providecommand \@@endlink[0]{}%
\providecommand \url  [0]{\begingroup\@sanitize@url \@url }%
\providecommand \@url [1]{\endgroup\@href {#1}{\urlprefix }}%
\providecommand \urlprefix  [0]{URL }%
\providecommand \Eprint [0]{\href }%
\providecommand \doibase [0]{http://dx.doi.org/}%
\providecommand \selectlanguage [0]{\@gobble}%
\providecommand \bibinfo  [0]{\@secondoftwo}%
\providecommand \bibfield  [0]{\@secondoftwo}%
\providecommand \translation [1]{[#1]}%
\providecommand \BibitemOpen [0]{}%
\providecommand \bibitemStop [0]{}%
\providecommand \bibitemNoStop [0]{.\EOS\space}%
\providecommand \EOS [0]{\spacefactor3000\relax}%
\providecommand \BibitemShut  [1]{\csname bibitem#1\endcsname}%
\let\auto@bib@innerbib\@empty
\bibitem [{\citenamefont {Crack}\ \emph {et~al.}(2014)\citenamefont {Crack},
  \citenamefont {Green}, \citenamefont {Thomson},\ and\ \citenamefont
  {Brun}}]{Crack_2014}%
  \BibitemOpen
  \bibfield  {author} {\bibinfo {author} {\bibfnamefont {J.~C.}\ \bibnamefont
  {Crack}}, \bibinfo {author} {\bibfnamefont {J.}~\bibnamefont {Green}},
  \bibinfo {author} {\bibfnamefont {A.~J.}\ \bibnamefont {Thomson}}, \ and\
  \bibinfo {author} {\bibfnamefont {N.~E.~L.}\ \bibnamefont {Brun}},\ }\href
  {\doibase 10.1021/ar5002507} {\bibfield  {journal} {\bibinfo  {journal}
  {Accounts of Chemical Research}\ }\textbf {\bibinfo {volume} {47}},\ \bibinfo
  {pages} {3196} (\bibinfo {year} {2014})}\BibitemShut {NoStop}%
\bibitem [{\citenamefont {Stiefel}(1996)}]{Stiefel_1996}%
  \BibitemOpen
  \bibfield  {author} {\bibinfo {author} {\bibfnamefont {E.~I.}\ \bibnamefont
  {Stiefel}},\ }\href {\doibase 10.1021/bk-1996-0653.ch001} {\bibfield
  {journal} {\bibinfo  {journal} {Transition Metal Sulfur Chemistry}\ ,\
  \bibinfo {pages} {2}} (\bibinfo {year} {1996})}\BibitemShut {NoStop}%
\bibitem [{\citenamefont {Xiao}\ \emph {et~al.}(2016)\citenamefont {Xiao},
  \citenamefont {Li}, \citenamefont {Sun}, \citenamefont {Guan},\ and\
  \citenamefont {Wang}}]{Xiao_2016}%
  \BibitemOpen
  \bibfield  {author} {\bibinfo {author} {\bibfnamefont {S.}~\bibnamefont
  {Xiao}}, \bibinfo {author} {\bibfnamefont {X.}~\bibnamefont {Li}}, \bibinfo
  {author} {\bibfnamefont {W.}~\bibnamefont {Sun}}, \bibinfo {author}
  {\bibfnamefont {B.}~\bibnamefont {Guan}}, \ and\ \bibinfo {author}
  {\bibfnamefont {Y.}~\bibnamefont {Wang}},\ }\href {\doibase
  10.1016/j.cej.2016.05.068} {\bibfield  {journal} {\bibinfo  {journal}
  {Chemical Engineering Journal}\ }\textbf {\bibinfo {volume} {306}},\ \bibinfo
  {pages} {251} (\bibinfo {year} {2016})}\BibitemShut {NoStop}%
\bibitem [{\citenamefont {Zhang}\ \emph {et~al.}(1996)\citenamefont {Zhang},
  \citenamefont {Hayase}, \citenamefont {Kawamata}, \citenamefont {Nakao},
  \citenamefont {Nakajima},\ and\ \citenamefont {Kaya}}]{Zhang_1996}%
  \BibitemOpen
  \bibfield  {author} {\bibinfo {author} {\bibfnamefont {N.}~\bibnamefont
  {Zhang}}, \bibinfo {author} {\bibfnamefont {T.}~\bibnamefont {Hayase}},
  \bibinfo {author} {\bibfnamefont {H.}~\bibnamefont {Kawamata}}, \bibinfo
  {author} {\bibfnamefont {K.}~\bibnamefont {Nakao}}, \bibinfo {author}
  {\bibfnamefont {A.}~\bibnamefont {Nakajima}}, \ and\ \bibinfo {author}
  {\bibfnamefont {K.}~\bibnamefont {Kaya}},\ }\href {\doibase 10.1063/1.471048}
  {\bibfield  {journal} {\bibinfo  {journal} {The Journal of Chemical Physics}\
  }\textbf {\bibinfo {volume} {104}},\ \bibinfo {pages} {3413} (\bibinfo {year}
  {1996})}\BibitemShut {NoStop}%
\bibitem [{\citenamefont {Takano}, \citenamefont {Yamamoto},\ and\
  \citenamefont {Saito}(2004)}]{Takano_2004}%
  \BibitemOpen
  \bibfield  {author} {\bibinfo {author} {\bibfnamefont {S.}~\bibnamefont
  {Takano}}, \bibinfo {author} {\bibfnamefont {S.}~\bibnamefont {Yamamoto}}, \
  and\ \bibinfo {author} {\bibfnamefont {S.}~\bibnamefont {Saito}},\ }\href
  {\doibase 10.1016/j.jms.2004.01.003} {\bibfield  {journal} {\bibinfo
  {journal} {Journal of Molecular Spectroscopy}\ }\textbf {\bibinfo {volume}
  {224}},\ \bibinfo {pages} {137} (\bibinfo {year} {2004})}\BibitemShut
  {NoStop}%
\bibitem [{\citenamefont {Drowart}, \citenamefont {Pattoret},\ and\
  \citenamefont {Smoes}(1967)}]{Drowart_1967}%
  \BibitemOpen
  \bibfield  {author} {\bibinfo {author} {\bibfnamefont {J.}~\bibnamefont
  {Drowart}}, \bibinfo {author} {\bibfnamefont {A.}~\bibnamefont {Pattoret}}, \
  and\ \bibinfo {author} {\bibfnamefont {S.}~\bibnamefont {Smoes}},\
  }\href@noop {} {\bibfield  {journal} {\bibinfo  {journal} {Proc. Br. Ceram.
  Soc.}\ }\textbf {\bibinfo {volume} {8}},\ \bibinfo {pages} {67?88}
  (\bibinfo {year} {1967})}\BibitemShut {NoStop}%
\bibitem [{\citenamefont {Wang}\ \emph {et~al.}(2011)\citenamefont {Wang},
  \citenamefont {Huang}, \citenamefont {Zhen}, \citenamefont {Zhang},\ and\
  \citenamefont {Chen}}]{Wang_2011}%
  \BibitemOpen
  \bibfield  {author} {\bibinfo {author} {\bibfnamefont {L.}~\bibnamefont
  {Wang}}, \bibinfo {author} {\bibfnamefont {D.-l.}\ \bibnamefont {Huang}},
  \bibinfo {author} {\bibfnamefont {J.-f.}\ \bibnamefont {Zhen}}, \bibinfo
  {author} {\bibfnamefont {Q.}~\bibnamefont {Zhang}}, \ and\ \bibinfo {author}
  {\bibfnamefont {Y.}~\bibnamefont {Chen}},\ }\href {\doibase
  10.1088/1674-0068/24/01/1-3} {\bibfield  {journal} {\bibinfo  {journal}
  {Chinese Journal of Chemical Physics}\ }\textbf {\bibinfo {volume} {24}},\
  \bibinfo {pages} {1} (\bibinfo {year} {2011})}\BibitemShut {NoStop}%
\bibitem [{\citenamefont {Matthew}, \citenamefont {Tieu},\ and\ \citenamefont
  {Morse}(2017)}]{Matthew_2017}%
  \BibitemOpen
  \bibfield  {author} {\bibinfo {author} {\bibfnamefont {D.~J.}\ \bibnamefont
  {Matthew}}, \bibinfo {author} {\bibfnamefont {E.}~\bibnamefont {Tieu}}, \
  and\ \bibinfo {author} {\bibfnamefont {M.~D.}\ \bibnamefont {Morse}},\ }\href
  {\doibase 10.1063/1.4979679} {\bibfield  {journal} {\bibinfo  {journal} {The
  Journal of Chemical Physics}\ }\textbf {\bibinfo {volume} {146}},\ \bibinfo
  {pages} {144310} (\bibinfo {year} {2017})}\BibitemShut {NoStop}%
\bibitem [{\citenamefont {Bridgeman}\ and\ \citenamefont
  {Rothery}(2000)}]{Bridgeman_2000}%
  \BibitemOpen
  \bibfield  {author} {\bibinfo {author} {\bibfnamefont {A.~J.}\ \bibnamefont
  {Bridgeman}}\ and\ \bibinfo {author} {\bibfnamefont {J.}~\bibnamefont
  {Rothery}},\ }\href {\doibase 10.1039/a906523g} {\bibfield  {journal}
  {\bibinfo  {journal} {Journal of the Chemical Society, Dalton Transactions}\
  ,\ \bibinfo {pages} {211}} (\bibinfo {year} {2000})}\BibitemShut {NoStop}%
\bibitem [{\citenamefont {Li}\ \emph {et~al.}(2013)\citenamefont {Li},
  \citenamefont {Wang}, \citenamefont {Wang}, \citenamefont {Gao},
  \citenamefont {Geng}, \citenamefont {Li}, \citenamefont {Wang},\ and\
  \citenamefont {Jiao}}]{Li_2013}%
  \BibitemOpen
  \bibfield  {author} {\bibinfo {author} {\bibfnamefont {Y.-N.}\ \bibnamefont
  {Li}}, \bibinfo {author} {\bibfnamefont {S.}~\bibnamefont {Wang}}, \bibinfo
  {author} {\bibfnamefont {T.}~\bibnamefont {Wang}}, \bibinfo {author}
  {\bibfnamefont {R.}~\bibnamefont {Gao}}, \bibinfo {author} {\bibfnamefont
  {C.-Y.}\ \bibnamefont {Geng}}, \bibinfo {author} {\bibfnamefont {Y.-W.}\
  \bibnamefont {Li}}, \bibinfo {author} {\bibfnamefont {J.}~\bibnamefont
  {Wang}}, \ and\ \bibinfo {author} {\bibfnamefont {H.}~\bibnamefont {Jiao}},\
  }\href {\doibase 10.1002/cphc.201201043} {\bibfield  {journal} {\bibinfo
  {journal} {ChemPhysChem}\ }\textbf {\bibinfo {volume} {14}},\ \bibinfo
  {pages} {1182} (\bibinfo {year} {2013})}\BibitemShut {NoStop}%
\bibitem [{\citenamefont {Liang}, \citenamefont {Wang},\ and\ \citenamefont
  {Andrews}(2009)}]{Liang_2009}%
  \BibitemOpen
  \bibfield  {author} {\bibinfo {author} {\bibfnamefont {B.}~\bibnamefont
  {Liang}}, \bibinfo {author} {\bibfnamefont {X.}~\bibnamefont {Wang}}, \ and\
  \bibinfo {author} {\bibfnamefont {L.}~\bibnamefont {Andrews}},\ }\href
  {\doibase 10.1021/jp900994c} {\bibfield  {journal} {\bibinfo  {journal} {The
  Journal of Physical Chemistry A}\ }\textbf {\bibinfo {volume} {113}},\
  \bibinfo {pages} {5375} (\bibinfo {year} {2009})}\BibitemShut {NoStop}%
\bibitem [{\citenamefont {Schultz}, \citenamefont {Zhao},\ and\ \citenamefont
  {Truhlar}(2005)}]{Schultz_2005}%
  \BibitemOpen
  \bibfield  {author} {\bibinfo {author} {\bibfnamefont {N.~E.}\ \bibnamefont
  {Schultz}}, \bibinfo {author} {\bibfnamefont {Y.}~\bibnamefont {Zhao}}, \
  and\ \bibinfo {author} {\bibfnamefont {D.~G.}\ \bibnamefont {Truhlar}},\
  }\href {\doibase 10.1021/jp0539223} {\bibfield  {journal} {\bibinfo
  {journal} {The Journal of Physical Chemistry A}\ }\textbf {\bibinfo {volume}
  {109}},\ \bibinfo {pages} {11127} (\bibinfo {year} {2005})}\BibitemShut
  {NoStop}%
\bibitem [{\citenamefont {Wu}, \citenamefont {Wang},\ and\ \citenamefont
  {Su}(2007)}]{Wu_2007}%
  \BibitemOpen
  \bibfield  {author} {\bibinfo {author} {\bibfnamefont {Z.~J.}\ \bibnamefont
  {Wu}}, \bibinfo {author} {\bibfnamefont {M.~Y.}\ \bibnamefont {Wang}}, \ and\
  \bibinfo {author} {\bibfnamefont {Z.~M.}\ \bibnamefont {Su}},\ }\href
  {\doibase 10.1002/jcc.20603} {\bibfield  {journal} {\bibinfo  {journal}
  {Journal of Computational Chemistry}\ }\textbf {\bibinfo {volume} {28}},\
  \bibinfo {pages} {703} (\bibinfo {year} {2007})}\BibitemShut {NoStop}%
\bibitem [{\citenamefont {H{\"u}bner}\ \emph {et~al.}(1998)\citenamefont
  {H{\"u}bner}, \citenamefont {Termath}, \citenamefont {Berning},\ and\
  \citenamefont {Sauer}}]{Hubner_1998}%
  \BibitemOpen
  \bibfield  {author} {\bibinfo {author} {\bibfnamefont {O.}~\bibnamefont
  {H{\"u}bner}}, \bibinfo {author} {\bibfnamefont {V.}~\bibnamefont {Termath}},
  \bibinfo {author} {\bibfnamefont {A.}~\bibnamefont {Berning}}, \ and\
  \bibinfo {author} {\bibfnamefont {J.}~\bibnamefont {Sauer}},\ }\href
  {\doibase 10.1016/s0009-2614(98)00792-1} {\bibfield  {journal} {\bibinfo
  {journal} {Chemical Physics Letters}\ }\textbf {\bibinfo {volume} {294}},\
  \bibinfo {pages} {37} (\bibinfo {year} {1998})}\BibitemShut {NoStop}%
\bibitem [{\citenamefont {Clima}\ and\ \citenamefont
  {Hendrickx}(2007)}]{Clima_2007}%
  \BibitemOpen
  \bibfield  {author} {\bibinfo {author} {\bibfnamefont {S.}~\bibnamefont
  {Clima}}\ and\ \bibinfo {author} {\bibfnamefont {M.~F.}\ \bibnamefont
  {Hendrickx}},\ }\href {\doibase 10.1016/j.cplett.2007.01.073} {\bibfield
  {journal} {\bibinfo  {journal} {Chemical Physics Letters}\ }\textbf {\bibinfo
  {volume} {436}},\ \bibinfo {pages} {341} (\bibinfo {year}
  {2007})}\BibitemShut {NoStop}%
\bibitem [{\citenamefont {Bauschlicher}\ and\ \citenamefont
  {Ma\^{i}tre}(1995)}]{Bauschlicher_1995}%
  \BibitemOpen
  \bibfield  {author} {\bibinfo {author} {\bibfnamefont {C.~W.}\ \bibnamefont
  {Bauschlicher}}\ and\ \bibinfo {author} {\bibfnamefont {P.}~\bibnamefont
  {Ma\^{i}tre}},\ }\href {\doibase 10.1007/s002140050068} {\bibfield  {journal}
  {\bibinfo  {journal} {Theoretica Chimica Acta}\ }\textbf {\bibinfo {volume}
  {90}},\ \bibinfo {pages} {189} (\bibinfo {year} {1995})}\BibitemShut
  {NoStop}%
\bibitem [{\citenamefont {Chen}\ \emph {et~al.}(2016)\citenamefont {Chen},
  \citenamefont {Zen}, \citenamefont {Brandenburg}, \citenamefont {Alf\`e},\
  and\ \citenamefont {Michaelides}}]{Chen_2016}%
  \BibitemOpen
  \bibfield  {author} {\bibinfo {author} {\bibfnamefont {J.}~\bibnamefont
  {Chen}}, \bibinfo {author} {\bibfnamefont {A.}~\bibnamefont {Zen}}, \bibinfo
  {author} {\bibfnamefont {J.~G.}\ \bibnamefont {Brandenburg}}, \bibinfo
  {author} {\bibfnamefont {D.}~\bibnamefont {Alf\`e}}, \ and\ \bibinfo {author}
  {\bibfnamefont {A.}~\bibnamefont {Michaelides}},\ }\href {\doibase
  10.1103/PhysRevB.94.220102} {\bibfield  {journal} {\bibinfo  {journal} {Phys.
  Rev. B}\ }\textbf {\bibinfo {volume} {94}},\ \bibinfo {pages} {220102}
  (\bibinfo {year} {2016})}\BibitemShut {NoStop}%
\bibitem [{\citenamefont {Dubeck{\'y}}, \citenamefont {Mit\'a\v{s}},\ and\
  \citenamefont {Jure{\v c}ka}(2016)}]{Dubeck__2016}%
  \BibitemOpen
  \bibfield  {author} {\bibinfo {author} {\bibfnamefont {M.}~\bibnamefont
  {Dubeck{\'y}}}, \bibinfo {author} {\bibfnamefont {L.}~\bibnamefont
  {Mit\'a\v{s}}}, \ and\ \bibinfo {author} {\bibfnamefont {P.}~\bibnamefont
  {Jure{\v c}ka}},\ }\href {\doibase 10.1021/acs.chemrev.5b00577} {\bibfield
  {journal} {\bibinfo  {journal} {Chemical Reviews}\ }\textbf {\bibinfo
  {volume} {116}},\ \bibinfo {pages} {5188} (\bibinfo {year}
  {2016})}\BibitemShut {NoStop}%
\bibitem [{\citenamefont {Zhou}\ and\ \citenamefont {Wang}(2017)}]{Zhou_2017}%
  \BibitemOpen
  \bibfield  {author} {\bibinfo {author} {\bibfnamefont {X.}~\bibnamefont
  {Zhou}}\ and\ \bibinfo {author} {\bibfnamefont {F.}~\bibnamefont {Wang}},\
  }\href {\doibase 10.1002/jcc.24750} {\bibfield  {journal} {\bibinfo
  {journal} {Journal of Computational Chemistry}\ }\textbf {\bibinfo {volume}
  {38}},\ \bibinfo {pages} {798} (\bibinfo {year} {2017})}\BibitemShut
  {NoStop}%
\bibitem [{\citenamefont {Guareschi}\ \emph {et~al.}(2016)\citenamefont
  {Guareschi}, \citenamefont {Zulfikri}, \citenamefont {Daday}, \citenamefont
  {Floris}, \citenamefont {Amovilli}, \citenamefont {Mennucci},\ and\
  \citenamefont {Filippi}}]{Filippi_2016}%
  \BibitemOpen
  \bibfield  {author} {\bibinfo {author} {\bibfnamefont {R.}~\bibnamefont
  {Guareschi}}, \bibinfo {author} {\bibfnamefont {H.}~\bibnamefont {Zulfikri}},
  \bibinfo {author} {\bibfnamefont {C.}~\bibnamefont {Daday}}, \bibinfo
  {author} {\bibfnamefont {F.~M.}\ \bibnamefont {Floris}}, \bibinfo {author}
  {\bibfnamefont {C.}~\bibnamefont {Amovilli}}, \bibinfo {author}
  {\bibfnamefont {B.}~\bibnamefont {Mennucci}}, \ and\ \bibinfo {author}
  {\bibfnamefont {C.}~\bibnamefont {Filippi}},\ }\href {\doibase
  10.1021/acs.jctc.6b00044} {\bibfield  {journal} {\bibinfo  {journal} {Journal
  of Chemical Theory and Computation}\ }\textbf {\bibinfo {volume} {12}},\
  \bibinfo {pages} {1674} (\bibinfo {year} {2016})},\ \bibinfo {note} {pMID:
  26959751},\ \Eprint
  {http://arxiv.org/abs/http://dx.doi.org/10.1021/acs.jctc.6b00044}
  {http://dx.doi.org/10.1021/acs.jctc.6b00044} \BibitemShut {NoStop}%
\bibitem [{\citenamefont {Christiansen}(1991)}]{Christiansen_1991}%
  \BibitemOpen
  \bibfield  {author} {\bibinfo {author} {\bibfnamefont {P.~A.}\ \bibnamefont
  {Christiansen}},\ }\href {\doibase 10.1063/1.461491} {\bibfield  {journal}
  {\bibinfo  {journal} {The Journal of Chemical Physics}\ }\textbf {\bibinfo
  {volume} {95}},\ \bibinfo {pages} {361} (\bibinfo {year} {1991})},\ \Eprint
  {http://arxiv.org/abs/https://doi.org/10.1063/1.461491}
  {https://doi.org/10.1063/1.461491} \BibitemShut {NoStop}%
\bibitem [{\citenamefont {Mit\'a\v{s}}(1993)}]{Mitas_1993}%
  \BibitemOpen
  \bibfield  {author} {\bibinfo {author} {\bibfnamefont {L.}~\bibnamefont
  {Mit\'a\v{s}}},\ }in\ \href@noop {} {\emph {\bibinfo {booktitle} {Computer
  Simulations Studies in Condensed Matter V}}},\ \bibinfo {editor} {edited by\
  \bibinfo {editor} {\bibfnamefont {D.~P.}\ \bibnamefont {Landau}}, \bibinfo
  {editor} {\bibfnamefont {K.~K.}\ \bibnamefont {Mon}}, \ and\ \bibinfo
  {editor} {\bibfnamefont {H.~B.}\ \bibnamefont {Sch\"uttler}}}\ (\bibinfo
  {publisher} {Springer, Berlin},\ \bibinfo {year} {1993})\ p.~\bibinfo {pages}
  {94}\BibitemShut {NoStop}%
\bibitem [{\citenamefont {Belohorec}, \citenamefont {Rothstein},\ and\
  \citenamefont {Vrbik}(1993)}]{Belohorec_93}%
  \BibitemOpen
  \bibfield  {author} {\bibinfo {author} {\bibfnamefont {P.}~\bibnamefont
  {Belohorec}}, \bibinfo {author} {\bibfnamefont {S.~M.}\ \bibnamefont
  {Rothstein}}, \ and\ \bibinfo {author} {\bibfnamefont {J.}~\bibnamefont
  {Vrbik}},\ }\href {\doibase 10.1063/1.464838} {\bibfield  {journal} {\bibinfo
   {journal} {The Journal of Chemical Physics}\ }\textbf {\bibinfo {volume}
  {98}},\ \bibinfo {pages} {6401} (\bibinfo {year} {1993})},\ \Eprint
  {http://arxiv.org/abs/https://doi.org/10.1063/1.464838}
  {https://doi.org/10.1063/1.464838} \BibitemShut {NoStop}%
\bibitem [{\citenamefont {Mit\'a\v{s}}(1994)}]{Mitas_1994}%
  \BibitemOpen
  \bibfield  {author} {\bibinfo {author} {\bibfnamefont {L.}~\bibnamefont
  {Mit\'a\v{s}}},\ }\href {\doibase 10.1103/PhysRevA.49.4411} {\bibfield
  {journal} {\bibinfo  {journal} {Phys. Rev. A}\ }\textbf {\bibinfo {volume}
  {49}},\ \bibinfo {pages} {4411} (\bibinfo {year} {1994})}\BibitemShut
  {NoStop}%
\bibitem [{\citenamefont {Sokolova}\ and\ \citenamefont
  {L{\"u}chow}(2000)}]{Sokolova_2000}%
  \BibitemOpen
  \bibfield  {author} {\bibinfo {author} {\bibfnamefont {S.}~\bibnamefont
  {Sokolova}}\ and\ \bibinfo {author} {\bibfnamefont {A.}~\bibnamefont
  {L{\"u}chow}},\ }\href {\doibase 10.1016/s0009-2614(00)00276-1} {\bibfield
  {journal} {\bibinfo  {journal} {Chemical Physics Letters}\ }\textbf {\bibinfo
  {volume} {320}},\ \bibinfo {pages} {421} (\bibinfo {year}
  {2000})}\BibitemShut {NoStop}%
\bibitem [{\citenamefont {Wagner}\ and\ \citenamefont
  {Mit\'a\v{s}}(2003)}]{Wagner_2003}%
  \BibitemOpen
  \bibfield  {author} {\bibinfo {author} {\bibfnamefont {L.}~\bibnamefont
  {Wagner}}\ and\ \bibinfo {author} {\bibfnamefont {L.}~\bibnamefont
  {Mit\'a\v{s}}},\ }\href {\doibase 10.1016/s0009-2614(03)00128-3} {\bibfield
  {journal} {\bibinfo  {journal} {Chemical Physics Letters}\ }\textbf {\bibinfo
  {volume} {370}},\ \bibinfo {pages} {412} (\bibinfo {year}
  {2003})}\BibitemShut {NoStop}%
\bibitem [{\citenamefont {Diedrich}, \citenamefont {L{\"u}chow},\ and\
  \citenamefont {Grimme}(2005)}]{Diedrich_2005}%
  \BibitemOpen
  \bibfield  {author} {\bibinfo {author} {\bibfnamefont {C.}~\bibnamefont
  {Diedrich}}, \bibinfo {author} {\bibfnamefont {A.}~\bibnamefont
  {L{\"u}chow}}, \ and\ \bibinfo {author} {\bibfnamefont {S.}~\bibnamefont
  {Grimme}},\ }\href {\doibase 10.1063/1.1846654} {\bibfield  {journal}
  {\bibinfo  {journal} {The Journal of Chemical Physics}\ }\textbf {\bibinfo
  {volume} {122}},\ \bibinfo {pages} {021101} (\bibinfo {year}
  {2005})}\BibitemShut {NoStop}%
\bibitem [{\citenamefont {Caffarel}\ \emph {et~al.}(2005)\citenamefont
  {Caffarel}, \citenamefont {Daudey}, \citenamefont {Heully},\ and\
  \citenamefont {Ram{\'\i}rez-Sol{\'\i}s}}]{Caffarel_2005}%
  \BibitemOpen
  \bibfield  {author} {\bibinfo {author} {\bibfnamefont {M.}~\bibnamefont
  {Caffarel}}, \bibinfo {author} {\bibfnamefont {J.-P.}\ \bibnamefont
  {Daudey}}, \bibinfo {author} {\bibfnamefont {J.-L.}\ \bibnamefont {Heully}},
  \ and\ \bibinfo {author} {\bibfnamefont {A.}~\bibnamefont
  {Ram{\'\i}rez-Sol{\'\i}s}},\ }\href {\doibase 10.1063/1.2011393} {\bibfield
  {journal} {\bibinfo  {journal} {The Journal of Chemical Physics}\ }\textbf
  {\bibinfo {volume} {123}},\ \bibinfo {pages} {094102} (\bibinfo {year}
  {2005})},\ \Eprint {http://arxiv.org/abs/https://doi.org/10.1063/1.2011393}
  {https://doi.org/10.1063/1.2011393} \BibitemShut {NoStop}%
\bibitem [{\citenamefont {Buend{\'\i}a}, \citenamefont {G{\'a}lvez},\ and\
  \citenamefont {Sarsa}(2006)}]{Buendia_2006}%
  \BibitemOpen
  \bibfield  {author} {\bibinfo {author} {\bibfnamefont {E.}~\bibnamefont
  {Buend{\'\i}a}}, \bibinfo {author} {\bibfnamefont {F.}~\bibnamefont
  {G{\'a}lvez}}, \ and\ \bibinfo {author} {\bibfnamefont {A.}~\bibnamefont
  {Sarsa}},\ }\href {\doibase https://doi.org/10.1016/j.cplett.2006.07.027}
  {\bibfield  {journal} {\bibinfo  {journal} {Chemical Physics Letters}\
  }\textbf {\bibinfo {volume} {428}},\ \bibinfo {pages} {241 } (\bibinfo {year}
  {2006})}\BibitemShut {NoStop}%
\bibitem [{\citenamefont {Wagner}\ and\ \citenamefont
  {Mit\'a\v{s}}(2007)}]{Wagner_2007}%
  \BibitemOpen
  \bibfield  {author} {\bibinfo {author} {\bibfnamefont {L.~K.}\ \bibnamefont
  {Wagner}}\ and\ \bibinfo {author} {\bibfnamefont {L.}~\bibnamefont
  {Mit\'a\v{s}}},\ }\href {\doibase 10.1063/1.2428294} {\bibfield  {journal}
  {\bibinfo  {journal} {The Journal of Chemical Physics}\ }\textbf {\bibinfo
  {volume} {126}},\ \bibinfo {pages} {034105} (\bibinfo {year}
  {2007})}\BibitemShut {NoStop}%
\bibitem [{\citenamefont {Bande}\ and\ \citenamefont
  {L{\"u}chow}(2008)}]{Bande_2008}%
  \BibitemOpen
  \bibfield  {author} {\bibinfo {author} {\bibfnamefont {A.}~\bibnamefont
  {Bande}}\ and\ \bibinfo {author} {\bibfnamefont {A.}~\bibnamefont
  {L{\"u}chow}},\ }\href {\doibase 10.1039/b803571g} {\bibfield  {journal}
  {\bibinfo  {journal} {Physical Chemistry Chemical Physics}\ }\textbf
  {\bibinfo {volume} {10}},\ \bibinfo {pages} {3371} (\bibinfo {year}
  {2008})}\BibitemShut {NoStop}%
\bibitem [{\citenamefont {Casula}\ \emph {et~al.}(2009)\citenamefont {Casula},
  \citenamefont {Marchi}, \citenamefont {Azadi},\ and\ \citenamefont
  {Sorella}}]{Casula_2009}%
  \BibitemOpen
  \bibfield  {author} {\bibinfo {author} {\bibfnamefont {M.}~\bibnamefont
  {Casula}}, \bibinfo {author} {\bibfnamefont {M.}~\bibnamefont {Marchi}},
  \bibinfo {author} {\bibfnamefont {S.}~\bibnamefont {Azadi}}, \ and\ \bibinfo
  {author} {\bibfnamefont {S.}~\bibnamefont {Sorella}},\ }\href {\doibase
  10.1016/j.cplett.2009.07.005} {\bibfield  {journal} {\bibinfo  {journal}
  {Chemical Physics Letters}\ }\textbf {\bibinfo {volume} {477}},\ \bibinfo
  {pages} {255} (\bibinfo {year} {2009})}\BibitemShut {NoStop}%
\bibitem [{\citenamefont {Bouab\c{c}a}, \citenamefont {Bra\"{\i}da},\ and\
  \citenamefont {Caffarel}(2010)}]{Caffarel_2010}%
  \BibitemOpen
  \bibfield  {author} {\bibinfo {author} {\bibfnamefont {T.}~\bibnamefont
  {Bouab\c{c}a}}, \bibinfo {author} {\bibfnamefont {B.}~\bibnamefont
  {Bra\"{\i}da}}, \ and\ \bibinfo {author} {\bibfnamefont {M.}~\bibnamefont
  {Caffarel}},\ }\href {\doibase 10.1063/1.3457364} {\bibfield  {journal}
  {\bibinfo  {journal} {The Journal of Chemical Physics}\ }\textbf {\bibinfo
  {volume} {133}},\ \bibinfo {pages} {044111} (\bibinfo {year} {2010})},\
  \Eprint {http://arxiv.org/abs/https://doi.org/10.1063/1.3457364}
  {https://doi.org/10.1063/1.3457364} \BibitemShut {NoStop}%
\bibitem [{\citenamefont {Petz}\ and\ \citenamefont
  {L{\"u}chow}(2011)}]{Petz_2011}%
  \BibitemOpen
  \bibfield  {author} {\bibinfo {author} {\bibfnamefont {R.}~\bibnamefont
  {Petz}}\ and\ \bibinfo {author} {\bibfnamefont {A.}~\bibnamefont
  {L{\"u}chow}},\ }\href {\doibase 10.1002/cphc.201000942} {\bibfield
  {journal} {\bibinfo  {journal} {ChemPhysChem}\ }\textbf {\bibinfo {volume}
  {12}},\ \bibinfo {pages} {2031} (\bibinfo {year} {2011})}\BibitemShut
  {NoStop}%
\bibitem [{\citenamefont {Buend{\'\i}a}\ \emph {et~al.}(2013)\citenamefont
  {Buend{\'\i}a}, \citenamefont {G{\'a}lvez}, \citenamefont {Maldonado},\ and\
  \citenamefont {Sarsa}}]{Buendia_2013}%
  \BibitemOpen
  \bibfield  {author} {\bibinfo {author} {\bibfnamefont {E.}~\bibnamefont
  {Buend{\'\i}a}}, \bibinfo {author} {\bibfnamefont {F.}~\bibnamefont
  {G{\'a}lvez}}, \bibinfo {author} {\bibfnamefont {P.}~\bibnamefont
  {Maldonado}}, \ and\ \bibinfo {author} {\bibfnamefont {A.}~\bibnamefont
  {Sarsa}},\ }\href {\doibase https://doi.org/10.1016/j.cplett.2012.12.055}
  {\bibfield  {journal} {\bibinfo  {journal} {Chemical Physics Letters}\
  }\textbf {\bibinfo {volume} {559}},\ \bibinfo {pages} {12 } (\bibinfo {year}
  {2013})}\BibitemShut {NoStop}%
\bibitem [{\citenamefont {Caffarel}\ \emph {et~al.}(2014)\citenamefont
  {Caffarel}, \citenamefont {Giner}, \citenamefont {Scemama},\ and\
  \citenamefont {Ram{\'\i}rez-Sol{\'\i}s}}]{Caffarel_2014}%
  \BibitemOpen
  \bibfield  {author} {\bibinfo {author} {\bibfnamefont {M.}~\bibnamefont
  {Caffarel}}, \bibinfo {author} {\bibfnamefont {E.}~\bibnamefont {Giner}},
  \bibinfo {author} {\bibfnamefont {A.}~\bibnamefont {Scemama}}, \ and\
  \bibinfo {author} {\bibfnamefont {A.}~\bibnamefont
  {Ram{\'\i}rez-Sol{\'\i}s}},\ }\href {\doibase 10.1021/ct5004252} {\bibfield
  {journal} {\bibinfo  {journal} {Journal of Chemical Theory and Computation}\
  }\textbf {\bibinfo {volume} {10}},\ \bibinfo {pages} {5286} (\bibinfo {year}
  {2014})}\BibitemShut {NoStop}%
\bibitem [{\citenamefont {Scemama}\ \emph {et~al.}(2014)\citenamefont
  {Scemama}, \citenamefont {Applencourt}, \citenamefont {Giner},\ and\
  \citenamefont {Caffarel}}]{Scemama_2014}%
  \BibitemOpen
  \bibfield  {author} {\bibinfo {author} {\bibfnamefont {A.}~\bibnamefont
  {Scemama}}, \bibinfo {author} {\bibfnamefont {T.}~\bibnamefont
  {Applencourt}}, \bibinfo {author} {\bibfnamefont {E.}~\bibnamefont {Giner}},
  \ and\ \bibinfo {author} {\bibfnamefont {M.}~\bibnamefont {Caffarel}},\
  }\href {\doibase 10.1063/1.4903985} {\bibfield  {journal} {\bibinfo
  {journal} {The Journal of Chemical Physics}\ }\textbf {\bibinfo {volume}
  {141}},\ \bibinfo {pages} {244110} (\bibinfo {year} {2014})}\BibitemShut
  {NoStop}%
\bibitem [{\citenamefont {Trail}\ and\ \citenamefont
  {Needs}(2015)}]{Trail_2015}%
  \BibitemOpen
  \bibfield  {author} {\bibinfo {author} {\bibfnamefont {J.~R.}\ \bibnamefont
  {Trail}}\ and\ \bibinfo {author} {\bibfnamefont {R.~J.}\ \bibnamefont
  {Needs}},\ }\href {\doibase 10.1063/1.4907589} {\bibfield  {journal}
  {\bibinfo  {journal} {The Journal of Chemical Physics}\ }\textbf {\bibinfo
  {volume} {142}},\ \bibinfo {pages} {064110} (\bibinfo {year} {2015})},\
  \Eprint {http://arxiv.org/abs/https://doi.org/10.1063/1.4907589}
  {https://doi.org/10.1063/1.4907589} \BibitemShut {NoStop}%
\bibitem [{\citenamefont {Doblhoff-Dier}\ \emph {et~al.}(2016)\citenamefont
  {Doblhoff-Dier}, \citenamefont {Meyer}, \citenamefont {Hoggan}, \citenamefont
  {Kroes},\ and\ \citenamefont {Wagner}}]{Doblhoff_Dier_2016}%
  \BibitemOpen
  \bibfield  {author} {\bibinfo {author} {\bibfnamefont {K.}~\bibnamefont
  {Doblhoff-Dier}}, \bibinfo {author} {\bibfnamefont {J.}~\bibnamefont
  {Meyer}}, \bibinfo {author} {\bibfnamefont {P.~E.}\ \bibnamefont {Hoggan}},
  \bibinfo {author} {\bibfnamefont {G.-J.}\ \bibnamefont {Kroes}}, \ and\
  \bibinfo {author} {\bibfnamefont {L.~K.}\ \bibnamefont {Wagner}},\ }\href
  {\doibase 10.1021/acs.jctc.6b00160} {\bibfield  {journal} {\bibinfo
  {journal} {Journal of Chemical Theory and Computation}\ }\textbf {\bibinfo
  {volume} {12}},\ \bibinfo {pages} {2583} (\bibinfo {year}
  {2016})}\BibitemShut {NoStop}%
\bibitem [{\citenamefont {Krogel}, \citenamefont {Santana},\ and\ \citenamefont
  {Reboredo}(2016)}]{Krogel_2016}%
  \BibitemOpen
  \bibfield  {author} {\bibinfo {author} {\bibfnamefont {J.~T.}\ \bibnamefont
  {Krogel}}, \bibinfo {author} {\bibfnamefont {J.~A.}\ \bibnamefont {Santana}},
  \ and\ \bibinfo {author} {\bibfnamefont {F.~A.}\ \bibnamefont {Reboredo}},\
  }\href {\doibase 10.1103/PhysRevB.93.075143} {\bibfield  {journal} {\bibinfo
  {journal} {Phys. Rev. B}\ }\textbf {\bibinfo {volume} {93}},\ \bibinfo
  {pages} {075143} (\bibinfo {year} {2016})}\BibitemShut {NoStop}%
\bibitem [{\citenamefont {Haghighi~Mood}\ and\ \citenamefont
  {L{\"u}chow}(2017)}]{Haghighi_Mood_2017}%
  \BibitemOpen
  \bibfield  {author} {\bibinfo {author} {\bibfnamefont {K.}~\bibnamefont
  {Haghighi~Mood}}\ and\ \bibinfo {author} {\bibfnamefont {A.}~\bibnamefont
  {L{\"u}chow}},\ }\href {\doibase 10.1021/acs.jpca.7b05798} {\bibfield
  {journal} {\bibinfo  {journal} {The Journal of Physical Chemistry A}\
  }\textbf {\bibinfo {volume} {121}},\ \bibinfo {pages} {6165} (\bibinfo {year}
  {2017})}\BibitemShut {NoStop}%
\bibitem [{\citenamefont {Giner}, \citenamefont {Scemama},\ and\ \citenamefont
  {Caffarel}(2013)}]{Giner_2013}%
  \BibitemOpen
  \bibfield  {author} {\bibinfo {author} {\bibfnamefont {E.}~\bibnamefont
  {Giner}}, \bibinfo {author} {\bibfnamefont {A.}~\bibnamefont {Scemama}}, \
  and\ \bibinfo {author} {\bibfnamefont {M.}~\bibnamefont {Caffarel}},\ }\href
  {\doibase 10.1139/cjc-2013-0017} {\bibfield  {journal} {\bibinfo  {journal}
  {Canadian Journal of Chemistry}\ }\textbf {\bibinfo {volume} {91}},\ \bibinfo
  {pages} {879} (\bibinfo {year} {2013})}\BibitemShut {NoStop}%
\bibitem [{\citenamefont {Giner}, \citenamefont {Scemama},\ and\ \citenamefont
  {Caffarel}(2015)}]{Giner_2015}%
  \BibitemOpen
  \bibfield  {author} {\bibinfo {author} {\bibfnamefont {E.}~\bibnamefont
  {Giner}}, \bibinfo {author} {\bibfnamefont {A.}~\bibnamefont {Scemama}}, \
  and\ \bibinfo {author} {\bibfnamefont {M.}~\bibnamefont {Caffarel}},\ }\href
  {\doibase 10.1063/1.4905528} {\bibfield  {journal} {\bibinfo  {journal} {The
  Journal of Chemical Physics}\ }\textbf {\bibinfo {volume} {142}},\ \bibinfo
  {pages} {044115} (\bibinfo {year} {2015})}\BibitemShut {NoStop}%
\bibitem [{\citenamefont {Scemama}\ \emph
  {et~al.}(2016{\natexlab{a}})\citenamefont {Scemama}, \citenamefont
  {Applencourt}, \citenamefont {Giner},\ and\ \citenamefont
  {Caffarel}}]{Scemama_2016}%
  \BibitemOpen
  \bibfield  {author} {\bibinfo {author} {\bibfnamefont {A.}~\bibnamefont
  {Scemama}}, \bibinfo {author} {\bibfnamefont {T.}~\bibnamefont
  {Applencourt}}, \bibinfo {author} {\bibfnamefont {E.}~\bibnamefont {Giner}},
  \ and\ \bibinfo {author} {\bibfnamefont {M.}~\bibnamefont {Caffarel}},\
  }\href {\doibase 10.1002/jcc.24382} {\bibfield  {journal} {\bibinfo
  {journal} {Journal of Computational Chemistry}\ }\textbf {\bibinfo {volume}
  {37}},\ \bibinfo {pages} {1866} (\bibinfo {year}
  {2016}{\natexlab{a}})}\BibitemShut {NoStop}%
\bibitem [{\citenamefont {Caffarel}\ \emph
  {et~al.}(2016{\natexlab{a}})\citenamefont {Caffarel}, \citenamefont
  {Applencourt}, \citenamefont {Giner},\ and\ \citenamefont
  {Scemama}}]{Caffarel_2016}%
  \BibitemOpen
  \bibfield  {author} {\bibinfo {author} {\bibfnamefont {M.}~\bibnamefont
  {Caffarel}}, \bibinfo {author} {\bibfnamefont {T.}~\bibnamefont
  {Applencourt}}, \bibinfo {author} {\bibfnamefont {E.}~\bibnamefont {Giner}},
  \ and\ \bibinfo {author} {\bibfnamefont {A.}~\bibnamefont {Scemama}},\ }\href
  {\doibase 10.1063/1.4947093} {\bibfield  {journal} {\bibinfo  {journal} {The
  Journal of Chemical Physics}\ }\textbf {\bibinfo {volume} {144}},\ \bibinfo
  {pages} {151103} (\bibinfo {year} {2016}{\natexlab{a}})}\BibitemShut
  {NoStop}%
\bibitem [{\citenamefont {Caffarel}\ \emph
  {et~al.}(2016{\natexlab{b}})\citenamefont {Caffarel}, \citenamefont
  {Applencourt}, \citenamefont {Giner},\ and\ \citenamefont
  {Scemama}}]{Caffarel_2016b}%
  \BibitemOpen
  \bibfield  {author} {\bibinfo {author} {\bibfnamefont {M.}~\bibnamefont
  {Caffarel}}, \bibinfo {author} {\bibfnamefont {T.}~\bibnamefont
  {Applencourt}}, \bibinfo {author} {\bibfnamefont {E.}~\bibnamefont {Giner}},
  \ and\ \bibinfo {author} {\bibfnamefont {A.}~\bibnamefont {Scemama}},\
  }\enquote {\bibinfo {title} {Using cipsi nodes in diffusion monte carlo},}\
  in\ \href {\doibase 10.1021/bk-2016-1234.ch002} {\emph {\bibinfo {booktitle}
  {Recent Progress in Quantum Monte Carlo}}}\ (\bibinfo {year} {2016})\
  Chap.~\bibinfo {chapter} {2}, pp.\ \bibinfo {pages} {15--46},\ \Eprint
  {http://arxiv.org/abs/http://pubs.acs.org/doi/pdf/10.1021/bk-2016-1234.ch002}
  {http://pubs.acs.org/doi/pdf/10.1021/bk-2016-1234.ch002} \BibitemShut
  {NoStop}%
\bibitem [{\citenamefont {Caffarel}\ \emph
  {et~al.}(2016{\natexlab{c}})\citenamefont {Caffarel}, \citenamefont
  {Applencourt}, \citenamefont {Giner},\ and\ \citenamefont {Scemama}}]{arxiv}%
  \BibitemOpen
  \bibfield  {author} {\bibinfo {author} {\bibfnamefont {M.}~\bibnamefont
  {Caffarel}}, \bibinfo {author} {\bibfnamefont {T.}~\bibnamefont
  {Applencourt}}, \bibinfo {author} {\bibfnamefont {E.}~\bibnamefont {Giner}},
  \ and\ \bibinfo {author} {\bibfnamefont {A.}~\bibnamefont {Scemama}},\ }\href
  {\doibase 10.1021/bk-2016-1234.ch002} {\  (\bibinfo {year}
  {2016}{\natexlab{c}}),\ 10.1021/bk-2016-1234.ch002},\ \Eprint
  {http://arxiv.org/abs/arXiv:1607.06742} {arXiv:1607.06742} \BibitemShut
  {NoStop}%
\bibitem [{\citenamefont {Umrigar}\ and\ \citenamefont
  {Filippi}(2005)}]{Umrigar_2005}%
  \BibitemOpen
  \bibfield  {author} {\bibinfo {author} {\bibfnamefont {C.~J.}\ \bibnamefont
  {Umrigar}}\ and\ \bibinfo {author} {\bibfnamefont {C.}~\bibnamefont
  {Filippi}},\ }\href {\doibase 10.1103/physrevlett.94.150201} {\bibfield
  {journal} {\bibinfo  {journal} {Physical Review Letters}\ }\textbf {\bibinfo
  {volume} {94}} (\bibinfo {year} {2005}),\
  10.1103/physrevlett.94.150201}\BibitemShut {NoStop}%
\bibitem [{\citenamefont {Toulouse}\ and\ \citenamefont
  {Umrigar}(2007)}]{Toulouse_2007}%
  \BibitemOpen
  \bibfield  {author} {\bibinfo {author} {\bibfnamefont {J.}~\bibnamefont
  {Toulouse}}\ and\ \bibinfo {author} {\bibfnamefont {C.~J.}\ \bibnamefont
  {Umrigar}},\ }\href {\doibase 10.1063/1.2437215} {\bibfield  {journal}
  {\bibinfo  {journal} {The Journal of Chemical Physics}\ }\textbf {\bibinfo
  {volume} {126}},\ \bibinfo {pages} {084102} (\bibinfo {year}
  {2007})}\BibitemShut {NoStop}%
\bibitem [{\citenamefont {Umrigar}\ \emph {et~al.}(2007)\citenamefont
  {Umrigar}, \citenamefont {Toulouse}, \citenamefont {Filippi}, \citenamefont
  {Sorella},\ and\ \citenamefont {Hennig}}]{Umrigar_2007}%
  \BibitemOpen
  \bibfield  {author} {\bibinfo {author} {\bibfnamefont {C.~J.}\ \bibnamefont
  {Umrigar}}, \bibinfo {author} {\bibfnamefont {J.}~\bibnamefont {Toulouse}},
  \bibinfo {author} {\bibfnamefont {C.}~\bibnamefont {Filippi}}, \bibinfo
  {author} {\bibfnamefont {S.}~\bibnamefont {Sorella}}, \ and\ \bibinfo
  {author} {\bibfnamefont {R.~G.}\ \bibnamefont {Hennig}},\ }\href {\doibase
  10.1103/physrevlett.98.110201} {\bibfield  {journal} {\bibinfo  {journal}
  {Physical Review Letters}\ }\textbf {\bibinfo {volume} {98}} (\bibinfo {year}
  {2007}),\ 10.1103/physrevlett.98.110201}\BibitemShut {NoStop}%
\bibitem [{\citenamefont {Toulouse}\ and\ \citenamefont
  {Umrigar}(2008)}]{Toulouse_2008}%
  \BibitemOpen
  \bibfield  {author} {\bibinfo {author} {\bibfnamefont {J.}~\bibnamefont
  {Toulouse}}\ and\ \bibinfo {author} {\bibfnamefont {C.~J.}\ \bibnamefont
  {Umrigar}},\ }\href {\doibase 10.1063/1.2908237} {\bibfield  {journal}
  {\bibinfo  {journal} {The Journal of Chemical Physics}\ }\textbf {\bibinfo
  {volume} {128}},\ \bibinfo {pages} {174101} (\bibinfo {year}
  {2008})}\BibitemShut {NoStop}%
\bibitem [{\citenamefont {Bender}\ and\ \citenamefont
  {Davidson}(1969)}]{Bender_1969}%
  \BibitemOpen
  \bibfield  {author} {\bibinfo {author} {\bibfnamefont {C.~F.}\ \bibnamefont
  {Bender}}\ and\ \bibinfo {author} {\bibfnamefont {E.~R.}\ \bibnamefont
  {Davidson}},\ }\href {\doibase 10.1103/physrev.183.23} {\bibfield  {journal}
  {\bibinfo  {journal} {Physical Review}\ }\textbf {\bibinfo {volume} {183}},\
  \bibinfo {pages} {23} (\bibinfo {year} {1969})}\BibitemShut {NoStop}%
\bibitem [{\citenamefont {Whitten}\ and\ \citenamefont
  {Hackmeyer}(1969)}]{Whitten_1969}%
  \BibitemOpen
  \bibfield  {author} {\bibinfo {author} {\bibfnamefont {J.~L.}\ \bibnamefont
  {Whitten}}\ and\ \bibinfo {author} {\bibfnamefont {M.}~\bibnamefont
  {Hackmeyer}},\ }\href {\doibase 10.1063/1.1671985} {\bibfield  {journal}
  {\bibinfo  {journal} {The Journal of Chemical Physics}\ }\textbf {\bibinfo
  {volume} {51}},\ \bibinfo {pages} {5584} (\bibinfo {year}
  {1969})}\BibitemShut {NoStop}%
\bibitem [{\citenamefont {Huron}, \citenamefont {Malrieu},\ and\ \citenamefont
  {Rancurel}(1973)}]{Huron_1973}%
  \BibitemOpen
  \bibfield  {author} {\bibinfo {author} {\bibfnamefont {B.}~\bibnamefont
  {Huron}}, \bibinfo {author} {\bibfnamefont {J.~P.}\ \bibnamefont {Malrieu}},
  \ and\ \bibinfo {author} {\bibfnamefont {P.}~\bibnamefont {Rancurel}},\
  }\href {\doibase 10.1063/1.1679199} {\bibfield  {journal} {\bibinfo
  {journal} {The Journal of Chemical Physics}\ }\textbf {\bibinfo {volume}
  {58}},\ \bibinfo {pages} {5745} (\bibinfo {year} {1973})}\BibitemShut
  {NoStop}%
\bibitem [{\citenamefont {Evangelisti}, \citenamefont {Daudey},\ and\
  \citenamefont {Malrieu}(1983)}]{Evangelisti_1983}%
  \BibitemOpen
  \bibfield  {author} {\bibinfo {author} {\bibfnamefont {S.}~\bibnamefont
  {Evangelisti}}, \bibinfo {author} {\bibfnamefont {J.-P.}\ \bibnamefont
  {Daudey}}, \ and\ \bibinfo {author} {\bibfnamefont {J.-P.}\ \bibnamefont
  {Malrieu}},\ }\href {\doibase 10.1016/0301-0104(83)85011-3} {\bibfield
  {journal} {\bibinfo  {journal} {Chemical Physics}\ }\textbf {\bibinfo
  {volume} {75}},\ \bibinfo {pages} {91} (\bibinfo {year} {1983})}\BibitemShut
  {NoStop}%
\bibitem [{\citenamefont {Cimiraglia}(1985)}]{Cimiraglia_1985}%
  \BibitemOpen
  \bibfield  {author} {\bibinfo {author} {\bibfnamefont {R.}~\bibnamefont
  {Cimiraglia}},\ }\href {\doibase 10.1063/1.449362} {\bibfield  {journal}
  {\bibinfo  {journal} {The Journal of Chemical Physics}\ }\textbf {\bibinfo
  {volume} {83}},\ \bibinfo {pages} {1746} (\bibinfo {year}
  {1985})}\BibitemShut {NoStop}%
\bibitem [{\citenamefont {Cimiraglia}\ and\ \citenamefont
  {Persico}(1987)}]{Cimiraglia_1987}%
  \BibitemOpen
  \bibfield  {author} {\bibinfo {author} {\bibfnamefont {R.}~\bibnamefont
  {Cimiraglia}}\ and\ \bibinfo {author} {\bibfnamefont {M.}~\bibnamefont
  {Persico}},\ }\href@noop {} {\bibfield  {journal} {\bibinfo  {journal}
  {Journal of computational chemistry}\ }\textbf {\bibinfo {volume} {8}},\
  \bibinfo {pages} {39} (\bibinfo {year} {1987})}\BibitemShut {NoStop}%
\bibitem [{\citenamefont {Illas}, \citenamefont {Rubio},\ and\ \citenamefont
  {Ricart}(1988)}]{Illas_1988}%
  \BibitemOpen
  \bibfield  {author} {\bibinfo {author} {\bibfnamefont {F.}~\bibnamefont
  {Illas}}, \bibinfo {author} {\bibfnamefont {J.}~\bibnamefont {Rubio}}, \ and\
  \bibinfo {author} {\bibfnamefont {J.~M.}\ \bibnamefont {Ricart}},\ }\href
  {\doibase 10.1063/1.455405} {\bibfield  {journal} {\bibinfo  {journal} {The
  Journal of Chemical Physics}\ }\textbf {\bibinfo {volume} {89}},\ \bibinfo
  {pages} {6376} (\bibinfo {year} {1988})}\BibitemShut {NoStop}%
\bibitem [{\citenamefont {Povill}, \citenamefont {Rubio},\ and\ \citenamefont
  {Illas}(1992)}]{Povill_1992}%
  \BibitemOpen
  \bibfield  {author} {\bibinfo {author} {\bibfnamefont {A.}~\bibnamefont
  {Povill}}, \bibinfo {author} {\bibfnamefont {J.}~\bibnamefont {Rubio}}, \
  and\ \bibinfo {author} {\bibfnamefont {F.}~\bibnamefont {Illas}},\
  }\href@noop {} {\bibfield  {journal} {\bibinfo  {journal} {Theoretical
  Chemistry Accounts: Theory, Computation, and Modeling (Theoretica Chimica
  Acta)}\ }\textbf {\bibinfo {volume} {82}},\ \bibinfo {pages} {229} (\bibinfo
  {year} {1992})}\BibitemShut {NoStop}%
\bibitem [{\citenamefont {Abrams}\ and\ \citenamefont
  {Sherrill}(2005)}]{Abrams_2005}%
  \BibitemOpen
  \bibfield  {author} {\bibinfo {author} {\bibfnamefont {M.~L.}\ \bibnamefont
  {Abrams}}\ and\ \bibinfo {author} {\bibfnamefont {C.~D.}\ \bibnamefont
  {Sherrill}},\ }\href {\doibase 10.1016/j.cplett.2005.06.107} {\bibfield
  {journal} {\bibinfo  {journal} {Chemical Physics Letters}\ }\textbf {\bibinfo
  {volume} {412}},\ \bibinfo {pages} {121} (\bibinfo {year}
  {2005})}\BibitemShut {NoStop}%
\bibitem [{\citenamefont {Bunge}\ and\ \citenamefont
  {Carb{\'o}-Dorca}(2006)}]{Bunge_2006}%
  \BibitemOpen
  \bibfield  {author} {\bibinfo {author} {\bibfnamefont {C.~F.}\ \bibnamefont
  {Bunge}}\ and\ \bibinfo {author} {\bibfnamefont {R.}~\bibnamefont
  {Carb{\'o}-Dorca}},\ }\href {\doibase 10.1063/1.2207621} {\bibfield
  {journal} {\bibinfo  {journal} {The Journal of Chemical Physics}\ }\textbf
  {\bibinfo {volume} {125}},\ \bibinfo {pages} {014108} (\bibinfo {year}
  {2006})}\BibitemShut {NoStop}%
\bibitem [{\citenamefont {Bytautas}\ and\ \citenamefont
  {Ruedenberg}(2009)}]{Bytautas_2009}%
  \BibitemOpen
  \bibfield  {author} {\bibinfo {author} {\bibfnamefont {L.}~\bibnamefont
  {Bytautas}}\ and\ \bibinfo {author} {\bibfnamefont {K.}~\bibnamefont
  {Ruedenberg}},\ }\href {\doibase 10.1016/j.chemphys.2008.11.021} {\bibfield
  {journal} {\bibinfo  {journal} {Chemical Physics}\ }\textbf {\bibinfo
  {volume} {356}},\ \bibinfo {pages} {64} (\bibinfo {year} {2009})}\BibitemShut
  {NoStop}%
\bibitem [{\citenamefont {Booth}, \citenamefont {Thom},\ and\ \citenamefont
  {Alavi}(2009)}]{Booth_2009}%
  \BibitemOpen
  \bibfield  {author} {\bibinfo {author} {\bibfnamefont {G.~H.}\ \bibnamefont
  {Booth}}, \bibinfo {author} {\bibfnamefont {A.~J.~W.}\ \bibnamefont {Thom}},
  \ and\ \bibinfo {author} {\bibfnamefont {A.}~\bibnamefont {Alavi}},\ }\href
  {\doibase 10.1063/1.3193710} {\bibfield  {journal} {\bibinfo  {journal} {The
  Journal of Chemical Physics}\ }\textbf {\bibinfo {volume} {131}},\ \bibinfo
  {pages} {054106} (\bibinfo {year} {2009})}\BibitemShut {NoStop}%
\bibitem [{\citenamefont {Knowles}(2015)}]{Knowles_2015}%
  \BibitemOpen
  \bibfield  {author} {\bibinfo {author} {\bibfnamefont {P.~J.}\ \bibnamefont
  {Knowles}},\ }\href {\doibase 10.1080/00268976.2014.1003621} {\bibfield
  {journal} {\bibinfo  {journal} {Molecular Physics}\ }\textbf {\bibinfo
  {volume} {113}},\ \bibinfo {pages} {1655} (\bibinfo {year}
  {2015})}\BibitemShut {NoStop}%
\bibitem [{\citenamefont {Garniron}\ \emph {et~al.}(2017)\citenamefont
  {Garniron}, \citenamefont {Scemama}, \citenamefont {Loos},\ and\
  \citenamefont {Caffarel}}]{Garniron_2017b}%
  \BibitemOpen
  \bibfield  {author} {\bibinfo {author} {\bibfnamefont {Y.}~\bibnamefont
  {Garniron}}, \bibinfo {author} {\bibfnamefont {A.}~\bibnamefont {Scemama}},
  \bibinfo {author} {\bibfnamefont {P.-F.}\ \bibnamefont {Loos}}, \ and\
  \bibinfo {author} {\bibfnamefont {M.}~\bibnamefont {Caffarel}},\ }\href
  {\doibase 10.1063/1.4992127} {\bibfield  {journal} {\bibinfo  {journal} {The
  Journal of Chemical Physics}\ }\textbf {\bibinfo {volume} {147}},\ \bibinfo
  {pages} {034101} (\bibinfo {year} {2017})}\BibitemShut {NoStop}%
\bibitem [{\citenamefont {Evangelista}(2014)}]{Evangelista_2014}%
  \BibitemOpen
  \bibfield  {author} {\bibinfo {author} {\bibfnamefont {F.~A.}\ \bibnamefont
  {Evangelista}},\ }\href {\doibase 10.1063/1.4869192} {\bibfield  {journal}
  {\bibinfo  {journal} {The Journal of Chemical Physics}\ }\textbf {\bibinfo
  {volume} {140}},\ \bibinfo {pages} {124114} (\bibinfo {year}
  {2014})}\BibitemShut {NoStop}%
\bibitem [{\citenamefont {Liu}\ and\ \citenamefont
  {Hoffmann}(2016)}]{Liu_2016}%
  \BibitemOpen
  \bibfield  {author} {\bibinfo {author} {\bibfnamefont {W.}~\bibnamefont
  {Liu}}\ and\ \bibinfo {author} {\bibfnamefont {M.~R.}\ \bibnamefont
  {Hoffmann}},\ }\href {\doibase 10.1021/acs.jctc.5b01099} {\bibfield
  {journal} {\bibinfo  {journal} {Journal of Chemical Theory and Computation}\
  }\textbf {\bibinfo {volume} {12}},\ \bibinfo {pages} {1169} (\bibinfo {year}
  {2016})}\BibitemShut {NoStop}%
\bibitem [{\citenamefont {Schriber}\ and\ \citenamefont
  {Evangelista}(2016)}]{Schriber_2016}%
  \BibitemOpen
  \bibfield  {author} {\bibinfo {author} {\bibfnamefont {J.~B.}\ \bibnamefont
  {Schriber}}\ and\ \bibinfo {author} {\bibfnamefont {F.~A.}\ \bibnamefont
  {Evangelista}},\ }\href {\doibase 10.1063/1.4948308} {\bibfield  {journal}
  {\bibinfo  {journal} {The Journal of Chemical Physics}\ }\textbf {\bibinfo
  {volume} {144}},\ \bibinfo {pages} {161106} (\bibinfo {year}
  {2016})}\BibitemShut {NoStop}%
\bibitem [{\citenamefont {Tubman}\ \emph {et~al.}(2016)\citenamefont {Tubman},
  \citenamefont {Lee}, \citenamefont {Takeshita}, \citenamefont {Head-Gordon},\
  and\ \citenamefont {Whaley}}]{Tubman_2016}%
  \BibitemOpen
  \bibfield  {author} {\bibinfo {author} {\bibfnamefont {N.~M.}\ \bibnamefont
  {Tubman}}, \bibinfo {author} {\bibfnamefont {J.}~\bibnamefont {Lee}},
  \bibinfo {author} {\bibfnamefont {T.~Y.}\ \bibnamefont {Takeshita}}, \bibinfo
  {author} {\bibfnamefont {M.}~\bibnamefont {Head-Gordon}}, \ and\ \bibinfo
  {author} {\bibfnamefont {K.~B.}\ \bibnamefont {Whaley}},\ }\href {\doibase
  10.1063/1.4955109} {\bibfield  {journal} {\bibinfo  {journal} {The Journal of
  Chemical Physics}\ }\textbf {\bibinfo {volume} {145}},\ \bibinfo {pages}
  {044112} (\bibinfo {year} {2016})}\BibitemShut {NoStop}%
\bibitem [{\citenamefont {Holmes}, \citenamefont {Tubman},\ and\ \citenamefont
  {Umrigar}(2016)}]{Holmes_2016}%
  \BibitemOpen
  \bibfield  {author} {\bibinfo {author} {\bibfnamefont {A.~A.}\ \bibnamefont
  {Holmes}}, \bibinfo {author} {\bibfnamefont {N.~M.}\ \bibnamefont {Tubman}},
  \ and\ \bibinfo {author} {\bibfnamefont {C.~J.}\ \bibnamefont {Umrigar}},\
  }\href {\doibase 10.1021/acs.jctc.6b00407} {\bibfield  {journal} {\bibinfo
  {journal} {Journal of Chemical Theory and Computation}\ }\textbf {\bibinfo
  {volume} {12}},\ \bibinfo {pages} {3674} (\bibinfo {year}
  {2016})}\BibitemShut {NoStop}%
\bibitem [{\citenamefont {Per}\ and\ \citenamefont {Cleland}(2017)}]{Per_2017}%
  \BibitemOpen
  \bibfield  {author} {\bibinfo {author} {\bibfnamefont {M.~C.}\ \bibnamefont
  {Per}}\ and\ \bibinfo {author} {\bibfnamefont {D.~M.}\ \bibnamefont
  {Cleland}},\ }\href {\doibase 10.1063/1.4981527} {\bibfield  {journal}
  {\bibinfo  {journal} {The Journal of Chemical Physics}\ }\textbf {\bibinfo
  {volume} {146}},\ \bibinfo {pages} {164101} (\bibinfo {year}
  {2017})}\BibitemShut {NoStop}%
\bibitem [{\citenamefont {Ohtsuka}\ and\ \citenamefont
  {Hasegawa}(2017)}]{Ohtsuka_2017}%
  \BibitemOpen
  \bibfield  {author} {\bibinfo {author} {\bibfnamefont {Y.}~\bibnamefont
  {Ohtsuka}}\ and\ \bibinfo {author} {\bibfnamefont {J.-y.}\ \bibnamefont
  {Hasegawa}},\ }\href {\doibase 10.1063/1.4993214} {\bibfield  {journal}
  {\bibinfo  {journal} {The Journal of Chemical Physics}\ }\textbf {\bibinfo
  {volume} {147}},\ \bibinfo {pages} {034102} (\bibinfo {year}
  {2017})}\BibitemShut {NoStop}%
\bibitem [{\citenamefont {Sharma}\ \emph {et~al.}(2017)\citenamefont {Sharma},
  \citenamefont {Holmes}, \citenamefont {Jeanmairet}, \citenamefont {Alavi},\
  and\ \citenamefont {Umrigar}}]{Sharma_2017}%
  \BibitemOpen
  \bibfield  {author} {\bibinfo {author} {\bibfnamefont {S.}~\bibnamefont
  {Sharma}}, \bibinfo {author} {\bibfnamefont {A.~A.}\ \bibnamefont {Holmes}},
  \bibinfo {author} {\bibfnamefont {G.}~\bibnamefont {Jeanmairet}}, \bibinfo
  {author} {\bibfnamefont {A.}~\bibnamefont {Alavi}}, \ and\ \bibinfo {author}
  {\bibfnamefont {C.~J.}\ \bibnamefont {Umrigar}},\ }\href {\doibase
  10.1021/acs.jctc.6b01028} {\bibfield  {journal} {\bibinfo  {journal} {Journal
  of Chemical Theory and Computation}\ }\textbf {\bibinfo {volume} {13}},\
  \bibinfo {pages} {1595} (\bibinfo {year} {2017})}\BibitemShut {NoStop}%
\bibitem [{\citenamefont {Holmes}, \citenamefont {Umrigar},\ and\ \citenamefont
  {Sharma}(2017)}]{Holmes_2017}%
  \BibitemOpen
  \bibfield  {author} {\bibinfo {author} {\bibfnamefont {A.~A.}\ \bibnamefont
  {Holmes}}, \bibinfo {author} {\bibfnamefont {C.~J.}\ \bibnamefont {Umrigar}},
  \ and\ \bibinfo {author} {\bibfnamefont {S.}~\bibnamefont {Sharma}},\ }\href
  {\doibase 10.1063/1.4998614} {\bibfield  {journal} {\bibinfo  {journal} {The
  Journal of Chemical Physics}\ }\textbf {\bibinfo {volume} {147}},\ \bibinfo
  {pages} {164111} (\bibinfo {year} {2017})}\BibitemShut {NoStop}%
\bibitem [{\citenamefont {Zimmerman}(2017)}]{Zimmerman_2017}%
  \BibitemOpen
  \bibfield  {author} {\bibinfo {author} {\bibfnamefont {P.~M.}\ \bibnamefont
  {Zimmerman}},\ }\href {\doibase 10.1063/1.4977727} {\bibfield  {journal}
  {\bibinfo  {journal} {The Journal of Chemical Physics}\ }\textbf {\bibinfo
  {volume} {146}},\ \bibinfo {pages} {104102} (\bibinfo {year}
  {2017})}\BibitemShut {NoStop}%
\bibitem [{\citenamefont {Scemama}\ \emph
  {et~al.}(2016{\natexlab{b}})\citenamefont {Scemama}, \citenamefont
  {Applencourt}, \citenamefont {Garniron}, \citenamefont {Giner}, \citenamefont
  {David},\ and\ \citenamefont {Caffarel}}]{QP}%
  \BibitemOpen
  \bibfield  {author} {\bibinfo {author} {\bibfnamefont {A.}~\bibnamefont
  {Scemama}}, \bibinfo {author} {\bibfnamefont {T.}~\bibnamefont
  {Applencourt}}, \bibinfo {author} {\bibfnamefont {Y.}~\bibnamefont
  {Garniron}}, \bibinfo {author} {\bibfnamefont {E.}~\bibnamefont {Giner}},
  \bibinfo {author} {\bibfnamefont {G.}~\bibnamefont {David}}, \ and\ \bibinfo
  {author} {\bibfnamefont {M.}~\bibnamefont {Caffarel}},\ }\href {\doibase
  10.5281/zenodo.200970} {\enquote {\bibinfo {title} {Quantum package v1.0},}\
  } (\bibinfo {year} {2016}{\natexlab{b}}),\ \bibinfo {note}
  {\url{https://github.com/LCPQ/quantum_package}}\BibitemShut {NoStop}%
\bibitem [{\citenamefont {Scemama}\ \emph {et~al.}(2017)\citenamefont
  {Scemama}, \citenamefont {Giner}, \citenamefont {Applencourt},\ and\
  \citenamefont {Caffarel}}]{qmcchem}%
  \BibitemOpen
  \bibfield  {author} {\bibinfo {author} {\bibfnamefont {A.}~\bibnamefont
  {Scemama}}, \bibinfo {author} {\bibfnamefont {E.}~\bibnamefont {Giner}},
  \bibinfo {author} {\bibfnamefont {T.}~\bibnamefont {Applencourt}}, \ and\
  \bibinfo {author} {\bibfnamefont {M.}~\bibnamefont {Caffarel}},\ }\href@noop
  {} {\enquote {\bibinfo {title} {Qmc=chem},}\ } (\bibinfo {year} {2017}),\
  \bibinfo {note} {https://github.com/scemama/qmcchem}\BibitemShut {NoStop}%
\bibitem [{\citenamefont {Scemama}\ \emph {et~al.}(2013)\citenamefont
  {Scemama}, \citenamefont {Caffarel}, \citenamefont {Oseret},\ and\
  \citenamefont {Jalby}}]{Scemama_2013}%
  \BibitemOpen
  \bibfield  {author} {\bibinfo {author} {\bibfnamefont {A.}~\bibnamefont
  {Scemama}}, \bibinfo {author} {\bibfnamefont {M.}~\bibnamefont {Caffarel}},
  \bibinfo {author} {\bibfnamefont {E.}~\bibnamefont {Oseret}}, \ and\ \bibinfo
  {author} {\bibfnamefont {W.}~\bibnamefont {Jalby}},\ }\href {\doibase
  10.1002/jcc.23216} {\bibfield  {journal} {\bibinfo  {journal} {Journal of
  Computational Chemistry}\ }\textbf {\bibinfo {volume} {34}},\ \bibinfo
  {pages} {938} (\bibinfo {year} {2013})}\BibitemShut {NoStop}%
\bibitem [{\citenamefont {Burkatzki}, \citenamefont {Filippi},\ and\
  \citenamefont {Dolg}(2007)}]{Burkatzki_2007}%
  \BibitemOpen
  \bibfield  {author} {\bibinfo {author} {\bibfnamefont {M.}~\bibnamefont
  {Burkatzki}}, \bibinfo {author} {\bibfnamefont {C.}~\bibnamefont {Filippi}},
  \ and\ \bibinfo {author} {\bibfnamefont {M.}~\bibnamefont {Dolg}},\ }\href
  {\doibase 10.1063/1.2741534} {\bibfield  {journal} {\bibinfo  {journal} {The
  Journal of Chemical Physics}\ }\textbf {\bibinfo {volume} {126}},\ \bibinfo
  {pages} {234105} (\bibinfo {year} {2007})}\BibitemShut {NoStop}%
\bibitem [{\citenamefont {Burkatzki}, \citenamefont {Filippi},\ and\
  \citenamefont {Dolg}(2008)}]{Burkatzki_2008}%
  \BibitemOpen
  \bibfield  {author} {\bibinfo {author} {\bibfnamefont {M.}~\bibnamefont
  {Burkatzki}}, \bibinfo {author} {\bibfnamefont {C.}~\bibnamefont {Filippi}},
  \ and\ \bibinfo {author} {\bibfnamefont {M.}~\bibnamefont {Dolg}},\ }\href
  {\doibase 10.1063/1.2987872} {\bibfield  {journal} {\bibinfo  {journal} {The
  Journal of Chemical Physics}\ }\textbf {\bibinfo {volume} {129}},\ \bibinfo
  {pages} {164115} (\bibinfo {year} {2008})}\BibitemShut {NoStop}%
\bibitem [{\citenamefont {Hammond}, \citenamefont {Reynolds},\ and\
  \citenamefont {Lester}(1987)}]{Hammond_1987}%
  \BibitemOpen
  \bibfield  {author} {\bibinfo {author} {\bibfnamefont {B.~L.}\ \bibnamefont
  {Hammond}}, \bibinfo {author} {\bibfnamefont {P.~J.}\ \bibnamefont
  {Reynolds}}, \ and\ \bibinfo {author} {\bibfnamefont {W.~A.}\ \bibnamefont
  {Lester}},\ }\href {\doibase 10.1063/1.453345} {\bibfield  {journal}
  {\bibinfo  {journal} {The Journal of Chemical Physics}\ }\textbf {\bibinfo
  {volume} {87}},\ \bibinfo {pages} {1130} (\bibinfo {year}
  {1987})}\BibitemShut {NoStop}%
\bibitem [{\citenamefont {Schmidt}\ \emph {et~al.}(1993)\citenamefont
  {Schmidt}, \citenamefont {Baldridge}, \citenamefont {Boatz}, \citenamefont
  {Elbert}, \citenamefont {Gordon}, \citenamefont {Jensen}, \citenamefont
  {Koseki}, \citenamefont {Matsunaga}, \citenamefont {Nguyen}, \citenamefont
  {Su},\ and\ \citenamefont {et~al.}}]{Schmidt_1993}%
  \BibitemOpen
  \bibfield  {author} {\bibinfo {author} {\bibfnamefont {M.~W.}\ \bibnamefont
  {Schmidt}}, \bibinfo {author} {\bibfnamefont {K.~K.}\ \bibnamefont
  {Baldridge}}, \bibinfo {author} {\bibfnamefont {J.~A.}\ \bibnamefont
  {Boatz}}, \bibinfo {author} {\bibfnamefont {S.~T.}\ \bibnamefont {Elbert}},
  \bibinfo {author} {\bibfnamefont {M.~S.}\ \bibnamefont {Gordon}}, \bibinfo
  {author} {\bibfnamefont {J.~H.}\ \bibnamefont {Jensen}}, \bibinfo {author}
  {\bibfnamefont {S.}~\bibnamefont {Koseki}}, \bibinfo {author} {\bibfnamefont
  {N.}~\bibnamefont {Matsunaga}}, \bibinfo {author} {\bibfnamefont {K.~A.}\
  \bibnamefont {Nguyen}}, \bibinfo {author} {\bibfnamefont {S.}~\bibnamefont
  {Su}}, \ and\ \bibinfo {author} {\bibnamefont {et~al.}},\ }\href {\doibase
  10.1002/jcc.540141112} {\bibfield  {journal} {\bibinfo  {journal} {Journal of
  Computational Chemistry}\ }\textbf {\bibinfo {volume} {14}},\ \bibinfo
  {pages} {1347} (\bibinfo {year} {1993})}\BibitemShut {NoStop}%
\bibitem [{\citenamefont {Assaraf}, \citenamefont {Caffarel},\ and\
  \citenamefont {Khelif}(2000)}]{Assaraf_2000}%
  \BibitemOpen
  \bibfield  {author} {\bibinfo {author} {\bibfnamefont {R.}~\bibnamefont
  {Assaraf}}, \bibinfo {author} {\bibfnamefont {M.}~\bibnamefont {Caffarel}}, \
  and\ \bibinfo {author} {\bibfnamefont {A.}~\bibnamefont {Khelif}},\ }\href
  {\doibase 10.1103/physreve.61.4566} {\bibfield  {journal} {\bibinfo
  {journal} {Physical Review E}\ }\textbf {\bibinfo {volume} {61}},\ \bibinfo
  {pages} {4566} (\bibinfo {year} {2000})}\BibitemShut {NoStop}%
\bibitem [{\citenamefont {Lee}\ \emph {et~al.}(2011)\citenamefont {Lee},
  \citenamefont {Conduit}, \citenamefont {Nemec}, \citenamefont
  {L{\'o}pez~R{\'\i}os},\ and\ \citenamefont {Drummond}}]{Lee_2011}%
  \BibitemOpen
  \bibfield  {author} {\bibinfo {author} {\bibfnamefont {R.~M.}\ \bibnamefont
  {Lee}}, \bibinfo {author} {\bibfnamefont {G.~J.}\ \bibnamefont {Conduit}},
  \bibinfo {author} {\bibfnamefont {N.}~\bibnamefont {Nemec}}, \bibinfo
  {author} {\bibfnamefont {P.}~\bibnamefont {L{\'o}pez~R{\'\i}os}}, \ and\
  \bibinfo {author} {\bibfnamefont {N.~D.}\ \bibnamefont {Drummond}},\ }\href
  {\doibase 10.1103/physreve.83.066706} {\bibfield  {journal} {\bibinfo
  {journal} {Physical Review E}\ }\textbf {\bibinfo {volume} {83}} (\bibinfo
  {year} {2011}),\ 10.1103/physreve.83.066706}\BibitemShut {NoStop}%
\bibitem [{Note1()}]{Note1}%
  \BibitemOpen
  \bibinfo {note} {The error bars have been obtained by fitting a large set of
  energy curves. Each of these curves is obtained from independent realizations
  of the statistical noise. Note that due to the absence of correlations in the
  statistical noise, the error bars obtained in this way are certainly
  overestimated.}\BibitemShut {Stop}%
\end{thebibliography}
\end{document}